 \documentclass{emulateapj}

\shorttitle{ALMACAL I: a deep and wide submm survey}
\shortauthors{Oteo et al.}

\usepackage{natbib}

\begin{document}


\title{ALMACAL I: First dual-band number counts from a deep and wide ALMA submm survey, free from cosmic variance}

\author{I.~Oteo\altaffilmark{1,2}, M.\,A.~Zwaan\altaffilmark{2},  R.\,J.~Ivison\altaffilmark{2,1}, I.~Smail\altaffilmark{3}, and A.\,D.~Biggs\altaffilmark{2}}

\affil{$^1$Institute for Astronomy, University of Edinburgh, Royal Observatory, Blackford Hill, Edinburgh EH9 3HJ UK}
\affil{$^2$European Southern Observatory, Karl-Schwarzschild-Str.\ 2, 85748 Garching, Germany}
\affil{$^3$Centre for Extragalactic Astronomy, Department of Physics, Durham University, South Road, Durham DH1 3LE UK}
\email{ivanoteogomez@gmail.com}

\begin{abstract}

We have exploited ALMA calibration observations to carry out a novel, wide and deep submm survey, {\sc almacal}.  These calibration data comprise a large number of observations of calibrator fields in a variety of frequency bands and array configurations.  Gathering together data acquired during multiple visits to many ALMA calibrators, it is possible to reach noise levels which allow the detection of faint dusty, star-forming galaxies (DSFGs) over a significant area.  In this paper we outline our survey strategy and report the first results.  We have analysed data for 69 calibrators, reaching depths of $\sim 25 \, {\rm \mu Jy \, beam^{-1}}$ at sub-arcsec resolution.  Adopting a conservative approach based on $\geq 5 \sigma$ detections, we have found eight and 11 DSFGs in ALMA bands 6 and 7, respectively, with flux densities $S_{\rm 1.2 mm} \geq 0.2 \, {\rm mJy}$.  The faintest galaxies would have been missed by even the deepest \emph{Herschel} surveys.  Our cumulative number counts have been determined independently at 870\,$\mu$m and 1.2\,mm, from a sparse sampling of the astronomical sky, and are thus relatively free of cosmic variance. The counts are lower than reported previously by a factor of at least $2\times$.   Future analyses will yield large, secure samples of DSFGs, with redshifts determined via detection of submm spectral lines. Uniquely, our strategy then allows morphological studies of very faint DSFGs -- representative of more normal star-forming galaxies than conventional submm galaxies (SMGs) -- in fields where self-calibration is feasible, yielding milliarcsecond spatial resolution.

\end{abstract}

\keywords{galaxy evolution; submm galaxies; dust emission; number counts}

%

\section{Introduction}\label{intro}

Submm surveys revolutionised the study of galaxy formation and evolution by uncovering a population of dusty starbursts with submm flux densities of a few mJy, the so-called submm galaxies \citep[SMGs,][]{Smail1997ApJ...490L...5S,Ivison1998MNRAS.298..583I,Barger1998Natur.394..248B,Hughes1998Natur.394..241H}.  These relatively bright SMGs, which have a median redshift of $z \sim 2.3$ and ${\rm SFR} \gtrsim 100 \, M_\odot \, {\rm yr}^{-1}$ \citep{Chapman2005}, have traditionally been found using bolometer cameras such as the Submmillimetre Common-User Bolometer Array \citep[SCUBA ---][]{1999MNRAS.303..659H} mounted on single-dish telescopes such as the 15-m James Clerk Maxwell Telescope \citep[e.g.][]{Coppin2006MNRAS.372.1621C,Weiss2009ApJ...707.1201W_LABOCA,Wardlow2011MNRAS.415.1479W,Geach2013MNRAS.432...53G,Casey2013MNRAS.436.1919C}. One of the main advantages of single-dish observations is that they can survey sufficiently wide areas of the sky to enable the detection of relatively large numbers of galaxies.  However, the large beam of single-dish telescopes makes it difficult to identify and explore the nature of even the brightest SMGs, unless dedicated interferometric follow-up observations are carried out, and any fainter population is buried in the confusion noise.

Given that significant numbers of submm surveys have led to the selection of large samples of bright SMGs, the next obvious step is to carry out blind searches using submm interferometers, detecting dusty star-forming galaxies (DSFGs) at sub-arcsec spatial resolution, at flux density levels below those accessible to single-dish telescopes. The main caveat to this approach is the small field-of-view (FoV) of interferometric observations at wavelengths probing the dust emission from star-forming galaxies, such that even covering areas as small as a few arcmin$^2$ is highly time intensive.

To overcome this impediment, we are taking advantage of ALMA calibration scans to carry out a submm survey.  Using calibrators it is possible to carry out a deep and wide submm survey with the necessary data coming `for free' from science projects dedicated to a wide variety of astrophysical topics.  A classical ALMA scheduling block (SB) comprises several steps, some of which involve observations of very bright, compact sources with submm flux densities of the order of a Jy to calibrate the amplitude and phase of the visibilities of the science targets, to set the flux density scale and/or measure the bandpass response.  Observations of such calibrators are essential and represent a significant fraction of each SB.  A long list of calibrators is used to calibrate ALMA science data\footnote{{\tt https://almascience.eso.org/sc/search}}.  Each calibrator will typically be observed several times, on different dates, in several different ALMA bands, as part of one or several SBs corresponding to one or several ALMA science projects.  By combining compatible data for a given calibrator we can reach r.m.s.\ noise levels sufficiently low to detect DSFGs within the primary beam (PB) centered on the calibrator.  As an example, a typical observation of a bandpass calibrator lasts about 5 minutes.  According to the ALMA sensitivity calculator, in that time it is possible to reach a continuum depth of about 60 ${\rm \mu Jy \, {beam}^{-1}}$ in ALMA band 6 with 36 antennas, sufficient to detect DSFGs in the vicinity of the bandpass calibrator with flux densities, $S_{\rm 1.2 \, mm} > 0.3 \, {\rm mJy}$ at 5$\sigma$, ignoring for the moment the possible effect of limited dynamic range caused by the presence of a very bright source in the middle of the map.  DSFGs are rare galaxies, even at sub-mJy flux densities, so despite reaching very low noise levels, data for many calibrators must be acquired to increase the area surveyed and allow the detection of a significant number. 

There are several, key advantages of using calibrators to look for and analyse high-redshift DSFGs.  ALMA calibrators tend to be observed in different projects with different ALMA configurations, ensuring excellent coverage of the $uv$ plane.  Perhaps more importantly, a number of calibrators will be observed at extremely high spatial resolution, if this is amongst the requirements of the science project within which they are observed.  The simultaneous presence in the PB of one or more DSFGs and a bright ALMA calibrator lends itself perfectly to self-calibration \citep[e.g.][see also \S\ref{section_extracting_data_from_archive}]{1984ARA&A..22...97P}, which permits accurate tracking of the complex gains and hence near-perfect imaging, even with the longest available interferometric baselines.  This enables us to analyse the morphological properties of any fortuitously located DSFGs in unprecedented detail.  The fact that each calibrator is often observed in several different ALMA frequency bands allows us to ensure that faint detections are genuine, with close to 100\% confidence (see \S\ref{section_source_selection}). Furthermore, such multi-band data allow us to study the spectral indices of DSFGs at matched spatial resolution, and sometimes to determine their redshifts via so-called `blind' detections of spectral lines such as CO.  Next, the exposure time necessary for follow-up observations of any DSFGs found using our approach are dramatically reduced, since the calibrator field is used for the science observations, meaning that only a brief scan of a flux-density standard is required by way of calibration.  Finally, the number counts we report, coming from a sparse sampling of the astronomical sky, are relatively free of cosmic variance \citep[see, for example,][]{2013MNRAS.428L...6S}.

As part of our ALMA submm survey we aim to: \textbf{(1)}: Study the submm number counts in different ALMA bands. So far, the number counts reported in previous work have been derived in different bands and conversion between flux densities has been required to provide a sufficient sample of galaxies. These conversion factors usually assume the classical FIR/submm SED of an SMG at $z \sim 2.3$ but not all submm detections are due to classical, dusty SMGs, and the redshift is rarely known for a faint source, meaning that these conversion factors between bands are uncertain, and consequently that multi-band information is the most robust way to derive number counts and carry out reliable comparisons with models of galaxy formation and evolution. \textbf{(2)}: Search for emission lines in the data cubes to determine the redshifts of the DSFGs via CO and other bright FIR/submm emission lines in the multi-band ALMA data \citep[e.g.][]{Weiss2013ApJ...767...88W}, and to constrain the CO luminosity function and the cosmic ${\rm H_2}$ density, similar to the studies carried out so far with PdBI \citep{Walter2014ApJ...782...79W,Decarli2014ApJ...782...78D} but over much larger areas and using deeper observations. \textbf{(3)}: Carry out a morphological analysis of the emission from DSFGs in multiple bands at matched spatial resolution.

In this work we focus on the description of the survey and the first derived number counts. Future work will describe the other aspects of the survey.

This paper is structured as follows: in \S\ref{section_extracting_data_from_archive} we explain how the ALMA data used in this paper were obtained, calibrated and combined. \S\ref{section_analysis} details how those data have been analysed to construct our sample of DSFGs, including analyses of the contamination from calibrator-related sources, spurious sources, survey completeness, effective area covered and the effect of flux boosting. \S\ref{section_number_counts} presents the number counts as determined at the current stage of the survey.  Our conclusions are summarised in \S \ref{concluuuuu}.

\section{Extracting and calibrating ALMA calibrator data}\label{section_extracting_data_from_archive}

In this section we explain how we retrieved and calibrated the public ALMA calibrator data and how we created the maps we then used to detect submm emitters.

We first used the ALMA calibrator archive to find all calibrators observed prior to 2015 July as part of science projects that passed quality control.  Since we are primarily interested in the selection of star-forming galaxies via the detection of their redshifted dust emission, we retrieved data in ALMA bands 6 (B6, around 250\,GHz or 1.2\,mm) and 7 (B7, around 345\,GHz or 870\,$\mu$m), where we then select sources (see \S\ref{section_source_selection}). The projects contributing to our sample represent a random selection of all the projects undertaken at ALMA, so the biases in terms of sky coverage are generated only by the latitude of ALMA, the annual weather patterns in the Atacama desert, and the positions of objects of interest to the astronomical community, e.g. the Milky Way.

For datasets outside the proprietary time period, we retrieved the full data deliveries from the ALMA archive.  For datasets that remained within the proprietary time, a ticket to the ALMA helpdesk was submitted with a request to obtain the part of the data that includes the calibrator scans.

During the execution of the ALMA calibration scripts, the bandpass calibrators are not always fully calibrated, in the sense that the calibration tables obtained from the so-called phase calibrators are always applied to the science targets but not always to the bandpass calibrators.  For this reason, we needed to recover the correct flux density scale of the bandpass calibrators from the flux tables in the data delivery packages and re-apply the calibration tables to the calibrators. 

Next, we created so-called `pseudo-continuum' measurement sets for which all channels in each spectral window were averaged.  These pseudo-continuum files were used to self-calibrate the calibrator data. Two rounds of self-calibration were applied, at first only in phase, then in both amplitude and phase, both with a solution interval equal to the integration time.  Instead of imaging the data inbetween the self-calibration steps, we used a point-source model that we fitted to the $uv$ data. The advantage of subtracting the point-source model for each observation separately is that any variability of the calibration source will not affect the calibration of the combined data.  For the majority of the datasets we found that these two rounds of self-calibration and $uv$-model fitting produced adequate results. Finally, the point-source model was subtracted from the visibility data. This procedure produced calibrated visibilities for the background region of each calibrator scan.  

The calibrator-subtracted visibilities for every SB were then imaged individually, without combining data for a given calibrator, using the {\sc clean} task. To do this, we defined a cleaning window of radius 1.5$\times$ the expected {\sc fwhm} of the PB (23$''$ and 17$''$ in B6 and B7, respectively) in each band and cleaned down to the r.m.s.\ of the dirty image.  We then inspected every map by eye,  discarding all datasets which showed evidence of poor calibration.  We also inspected the calibrated visibilities to discard any poorly-calibrated SBs. This led to the loss of a significant number of datasets ($\approx 20$\%), but their size and complexity made flagging and re-calibration impracticable.

\begin{table*}
\begin{center}
\caption{Summary of the observations at the present stage of the survey}\label{hola_tabla}
\begin{tabular}{l c c c c c c c c c c}
\hline
Calibrator 	& R.A. & Dec. & $z$	& $\sigma_{\rm B6}$ & $\theta_{\rm B6}$ & $\sigma_{\rm B7}$ & $\theta_{\rm B7}$  \\
		&	[J2000] & [J2000] & & [$\mu$Jy beam$^{-1}$] & [$''$] & [$\mu$Jy beam$^{-1}$] & [$''$]     \\
\hline
\hline
J0011$-$2612	&	00:11:01.2		&	$-$26:12:33.1	&	1.096	&	47	&	$0.35 \times 0.30$	&				&					\\
J0121+1149	&	01:21:41.6	&	+11:49:50.4	&	0.570	&	85	&	$0.52 \times 0.50$	&				&					\\
J0132$-$1654	&	01:32:43.5	&	$-$16:54:48.5	&	1.020	&	84	&	$0.45 \times 0.34$	&				&					\\
J0141$-$0928	&	01:41:25.8	&	$-$09:28:43.7	&	0.733	&	98	&	$0.37 \times 0.22$	&				&					\\	
J0145$-$2733	&	01:45:03.4	&	$-$27:33:34.3	&	1.155	&	91	&	$0.28 \times 0.23$	&				&					\\
J0238+1636	&	02:38:38.9	&	+16:36:60.1	&	0.940	&	47	&	$0.41 \times 0.35$	&				&					\\
J0239$-$0234	&	02:39:45.5 	&	$-$02:34:41.0 	&	1.116	&	75	&	$0.55 \times 0.51$	&				&					\\
J0329$-$2357	&	03:29:54.1	&	$-$23:57:08.8	&	0.895	&	73	&	$0.31 \times 0.22$	&				&					\\	
J0519$-$4546	&	05:19:49.7	&	$-$45:46:43.9	&	0.035	&	39	&	$0.33 \times 0.24$	&				&					\\	
J0550$-$5732	&	05:50:09.6	&	$-$57:32:24.4	&	2.001	&	93	&	$0.37 \times 0.24$	&				&					\\	
J0854+2006	&	08:54:48.9	&	+20:06:30.6	&	0.306	&	53	&	$0.73 \times 0.69$	&				&					\\	
J1215+1654	&	12:15:04.0	&	+16:54:38.0	&	1.132	&	92	&	$1.43 \times 0.93$	&				&					\\
J1308$-$6707	&	13:08:17.2	&	$-$67:07:05.0	&			&	93	&	$0.33 \times 0.27$	&				&					\\
J1342$-$2900	&	13:42:15.3	&	$-$29:00:41.8	&	1.442	&	87	&	$0.68 \times 0.51$	&				&					\\	
J1550+0527	&	15:50:35.3	&	+05:27:10.5	&	1.422	&	39	&	$0.63 \times 0.45$	&				&					\\	
J1826$-$2924	&	18:26:20.6	&	$-$29:24:25.0	&			&	67	&	$0.53 \times 0.43$	&				&					\\
J1832$-$1035 &	18:32:20.8	&	$-$10:35:11.2	&			&	37	&	$0.50 \times 0.37$	&				&					\\
J1933$-$6942	&	19:33:31.2	&	$-$69:42:58.9	&	1.481	&	96	&	$0.85 \times 0.50$	&				&					\\
J1955+1358	&	19:55:11.6		&	+13:58:16.2	&	0.743	&	88	&	$0.61 \times 0.51$	&				&					\\		
J2009$-$4849	&	20:09:25.4	&	$-$48:49:53.7	&	0.071	&	49	&	$0.52 \times 0.43$	&				&					\\
J2223$-$3137	&	22:23:21.6	&	$-$31:37:02.1	&			&	37	&	$0.59 \times 0.52$	&				&					\\
 J2306$-$0459&	23:06:15.3       &	$-$04:59:48.3	&	1.139	&	67	&	$0.50 \times 0.36$	&				&					\\
J2331$-$1556	&	23:31:38.6	&	$-$15:56:57.2	&	1.153	&	32	&	$0.43 \times 0.31$	&				&					\\
J2357$-$5311	&	23:57:53.2	&	$-$53:11:14.0	&	1.006	&	78	&	$0.64 \times 0.39$	&				&					\\


\hline

J0217+0144	& 	02:17:49.0	&	+01:44:49.7	&	1.715	&				&					&	44	&	$0.48 \times 0.34$	\\
J0339$-$0146	&	03:39:30.9	&	$-$01:46:35.8	&	0.852	&				&					&	92	&	$0.35 \times 0.30$	\\
J0607$-$0834	& 	06:07:59.7	&	$-$08:34:50.0	&	0.872	&				&					&	41	&	$0.34 \times 0.30$	\\
J1048$-$1909	& 	10:48:06.6	&	 $-$19:09:35.7&	0.595	&				&					&	65	&	$1.27 \times 0.58$	\\
J1303$-$5540	& 	13:03:49.2	&	$-$55:40:31.6	&			&				&					&	57	&	$0.37 \times 0.29$	\\
J1347+1217	& 	13:47:33.4	& 	+12:17:24.2	&	0.122	&				&					&	40	&	$0.61 \times 0.42$	 \\ 
J1505+0326	& 	15:05:06.5	& 	+03:26:30.8	&	0.408	&				&					&	93	&	$0.46 \times 0.29$	\\
J1625$-$2527	& 	16:25:46.9	& 	$-$25:27:38.3	&	0.786	&				&					&	74	&	$0.52 \times 0.29$	\\
J1700$-$2610	& 	17:00:53.2	& 	$-$26:10:51.7	&			&				&					&	97	&	$0.64 \times 0.41$	\\
J1744$-$3116	& 	17:44:23.6	& 	$-$31:16:36.3	&			&				&					&	60	&	$0.33 \times 0.29$	\\
J1924$-$2914	& 	19:24:51.1	& 	$-$29:14:30.1	&	0.353	&				&					&	30	&	$0.33 \times 0.29$	\\
J2206$-$0031	& 	22:06:43.3	& 	$-$00:31:02.5	&	0.335	&				&					&	48	&	$0.36 \times 0.31$	\\	
J2232+1143 	& 	22:32:36.4	& 	+11:43:50.9	&	1.037	&				&					&	41	&	$0.41 \times 0.37$	\\


\hline

J0038$-$2459	&	00:38:14.7	&	$-$24:59:02.5	&	0.498	&	86		&	$0.49 \times 0.39$	&	231		&	$0.43 \times 0.29$	\\
J0108+0135	&	01:08:38.8	&	+01:35:00.8	&	2.099	&	62		&	$0.69 \times 0.49$	&	129		&	$1.02 \times 0.56$	\\
J0215$-$0222 &	02:15:42.0	&	$-$02:22:56.8	&	1.178	&	58		&	$0.65 \times 0.52$	&	32		&	$0.39 \times 0.34$	\\
J0217+0144	&	02:17:48.9	&	+01:44:50.0	&	1.715	&	52		&	$0.31 \times 0.25$	&	54		&	$0.33 \times 0.27$   	\\
J0224+0659	&	02:24:28.4	&	+06:59:23.3	&	0.511	&	75		&	$0.72 \times 0.59$	&	114		&	$0.36 \times 0.31$	\\	
J0241$-$0815	&	02:41:04.8	&	$-$08:15:20.8	&	0.005	&	28		&	$0.54 \times 0.37$	&	36		&	$0.41 \times 0.36$	\\
J0334$-$4008	&	03:34:13.7	&	$-$40:08:25.4	&	1.445	&	35		&	$0.34 \times 0.25$	&	72		&	$0.29 \times 0.21$	\\
J0348$-$2749	&	03:48:38.1	&	$-$27:49:13.6	&	0.991	&	85		&	$0.37 \times 0.23$	&	59		&	$0.28 \times 0.22$	\\	
J0423$-$0120	&	04:23:15.8	&	$-$01:20:33.1	&	0.916	&	29		&	$0.46 \times 0.44$	&	61		&	$0.33 \times 0.23$	\\
J0510+1800	&	05:10:02.4	&	+18:00:41.6	&	0.416	&	33		&	$0.54 \times 0.42$	&	65		&	$0.32 \times 0.23$	\\
J0522$-$3627	&	05:22:58.0 	&	$-$36:27:31.0 	&	0.057	&	60		&	$0.74 \times 0.55$	&	123		&	$0.44 \times 0.28$	\\
J0538$-$4405	&	05:38:50.3	&	$-$44:05:08.9	&	0.890	&	46		&	$0.42 \times 0.32$	&	62		&	$0.61 \times 0.46$	\\
J0635$-$7516	&	06:35:46.5	&	$-$75:16:16.8	&	0.653	&	23		&	$0.37 \times 0.28$	&	48		&	$0.32 \times 0.21$	\\	
J0825+0309	&	08:25:50.3	&	+03:09:24.5	&	0.506	&	39		&	$0.50 \times 0.42$	&	56		&	$1.05 \times 0.58$	\\	
J0909+0121	&	09:09:10.1	&	+01:21:35.6	&	1.025	&	68		&	$1.28\times 0.89$	&	48		&	$1.14 \times 0.55$	\\	
J1008+0621	&	10:08:00.8	&	+06:21:21.2	&	1.720	&	30		&	$0.50 \times 0.48$	&	51		&	$0.28 \times 0.23$	\\ 
J1010$-$0200 &	10:10:51.7	&	$-$02:00:19.6	&	0.890	&	29		&	$1.25 \times 0.73$	&	39		&	$0.91 \times 0.51$	\\	
J1037$-$2934	&	10:37:16.1	&	$-$29:34:02.8	&	0.312	&	45		&	$1.30 \times 0.70$	&	93		&	$0.80 \times 0.50$	\\
J1058+0133	&	10:58:29.6	&	+01:33:58.8	&	0.890	&	44		&	$0.50 \times 0.47$	&	82		&	$0.31 \times 0.23$	\\
J1215$-$1731	&	12:15:46.8	&	$-$17:31:45.4	&			&	83		&	$0.94 \times 0.43$	&	54		&	$0.38 \times 0.28$	\\
J1229+0203	&	12:29:06.7	&	+02:03:08.6	&	0.158	&	171		&	$0.90 \times 0.68$	&	80		& 	$0.62 \times 0.43$ 	\\
J1337$-$1257	&	13:37:39.8	&	$-$12:57:24.7	&	0.539	&	83		&	$0.71 \times 0.52$	&	50		&	$0.56 \times 0.43$	\\
J1427$-$4206	&	14:27:56.3	&	$-$42:06:19.4	&	1.522	&	35		&	$0.57 \times 0.47$	&	43		&	$0.38 \times 0.32$	\\
J1517$-$2422	&	15:17:41.8	&	$-$24:22:19.5	&	0.049	&	93		&	$0.69 \times 0.30$	&	42 		&	$0.38 \times 0.32$	\\
J1534$-$3526 &	15:34:54.7	&	$-$35:26:23.0	&	1.515	&	50		&	$0.51 \times 0.43$	&	116		&	$0.46 \times 0.35$	\\
J1617$-$5848	&	16:17:17.9	&	$-$58:48:07.9	&			&	51		&	$0.38 \times 0.24$	&	154		&	$0.46 \times 0.37$	\\	
J1733$-$1304	&	17:33:02.7	&	$-$13:04:50.0	&	0.902	&	45		&	$1.06 \times 0.55$	&	42		&	$0.39 \times 0.36$	\\
 J1832$-$2039	&	18:32:11.0		&	$-$20:39:48.2	&     0.103        &	87		&	$0.43 \times 0.38$	&	50		&	$0.40 \times 0.33$	\\
J2056$-$4714	&	20:56:16.3	&	$-$47:14:48.5	&	1.489	&	26		&	$0.47 \times 0.44$	&	41		&	$0.56 \times 0.47$	\\
J2148+0657	&	21:48:05.5	&	+06:57:38.6	&	0.791	&	34 		&	$0.51 \times 0.46$	&	56		&	$0.31 \times 0.23$	\\
CTA102		&	22:32:36.4	&	+11:43:50.8	&	1.037	&	35		&	$0.55 \times 0.42$	&	34		&	$0.55 \times 0.42$	\\
J2258$-$2758	&	22:58:06.0	&	$-$27:58:21.2	&	0.926	&	25		&	$0.36 \times 0.30$	&	94		&	$0.38 \times 0.31$	\\


\hline
\end{tabular}
\end{center}
\end{table*}

Having discarded all bad datasets, we re-calculated the weights of the visibilities of the remaining datasets using {\sc statwt} to measure the visibility scatter empirically as a function of time, antenna, and/or baseline for each dataset, re-weighting the visibilities accordingly, before combining all data for each calibrator in each band with {\sc concat}.  We then created the deep maps following the same cleaning technique explained earlier.  Table \ref{hola_tabla} provides a summary of the calibrators used in this survey, along with the depth and beam size of each individual map.  We have determined the {\sc fwhm} of each calibrator map by fitting a Gaussian profile to the PB response obtained during the cleaning process in {\sc casa}. We have used 69 pointings, covering a total area close to 19\,arcmin$^2$ for bright DSFGs, $S_{\rm 870 \mu m} > 1\,{\rm mJy}$, although the effective area for a given flux density depends on the depth of the maps -- see \S\ref{section_area_covered_by_the_survey}. It can be seen in Table~\ref{hola_tabla} that most of our ALMA maps have sub-arcsec resolution and reach r.m.s.\ noise levels, $\sigma \sim 30\mu$Jy\,beam$^{-1}$, in B6 and B7.

We anticipated that one possible limitation of our approach might be the presence of the bright calibrator in the middle of the map, which may have been expected to influence the dynamic range of the image, defined as the ratio between the flux density of the brightest source detected and the r.m.s.\ of the map.  However, the r.m.s.\ of the clean maps listed in Table~\ref{hola_tabla} indicates that we have reached dynamic ranges in excess of 18,000, detecting DSFGs as faint as $S_{\rm 1.2 mm} \sim 0.2 \, {\rm mJy}$ at the present stage of the survey (see \S\ref{section_source_catalog}).  We conclude, therefore, that the presence of the calibrator in the middle of the image has not proved to be a strong limitation.

\section{Analysis}\label{section_analysis}

\subsection{Source detection}\label{section_source_selection}

As has been the case with most previous reports of faint ALMA counts \citep[e.g.][]{Hatsukade2013ApJ...769L..27H,Ono2014ApJ...795....5O,Fujimoto2015arXiv150503523F,Carniani2015arXiv150200640C}, our source detection has been performed using SExtractor \citep{Bertin1996} on the clean maps, before correcting for the antenuation response of the PB.  For those calibrators for which we have both B6 and B7 data, we searched for sources in B6 because the maps have a larger FoV.  Searching for sources in B7 provided no additional sources.  Because calibrators are such bright sources, we might expect some low-level residual emission in the maps, even after subtracting the calibrators from the data using a point-source model.  Faced with this fear, we adopted a conservative source detection technique to exclude the possibility of spurious detections: we selected sources with peak flux densities at least 5$\times$ the r.m.s.\ noise, rather higher than the ${\rm SNR \sim 4}$ limit used in previous work.  Our higher SNR threshold means we do not detect very faint sources, but we reduce the number of false detections considerably. Despite our conservative approach, we were able to detect sources down to $S_{\rm 870 \mu m} \sim 0.4 \, {\rm mJy}$ at sub-arcsec spatial resolution (see \S\ref{section_source_catalog}), even at the present stage of the survey.

After correcting for the PB response, the r.m.s.\ of the clean maps increases rapidly with the distance from the map center and hence our sensitivity to sources at large distances from the map center decreases. We searched for detections within a radius of 1.5$\times$ the {\sc fwhm} of the PB of each map, a region which is not severely affected by PB attenuation. A $\approx 1$-mJy galaxy was detected close to the edge of the detection area around the calibrator, J1744$-$3116, showing that this detection area does not include inaccessible parts of the PB-corrected maps.

Whenever we detected an DSFG around a calibrator, the ALMA archive was queried again, this time with no limitation placed on the ALMA observing band.  This additional search is done to: \textbf{(1)} increase the SNR of the detection; \textbf{(2)} confirm via multi-band observations that faint sources are real; \textbf{(3)} look for emission lines in the FIR/submm spectrum of the submm emitter to determine or constrain its redshift; \textbf{(4)} exploit multi-band information to distinguish between genuine DSFGs and jets emanating from the calibrator or other calibrator-related emission.  Data obtained in ALMA bands 3 and 4 often allow the detection of the brightest mid-$J$ CO lines  \citep{2007ASPC..375...25W,2009ApJ...705L..45W, 2013ApJ...767...88W} and also represent the most efficient way to distinguish between synchrotron-powered jets and thermal emission from DSFGs; data obtained in ALMA bands 8 and 9 can improve our sampling of the FIR/submm SED -- at least for sources close to the calibrator, within the smaller, high-frequency FoV.

\begin{figure*}
\centering
\includegraphics[width=0.19\textwidth]{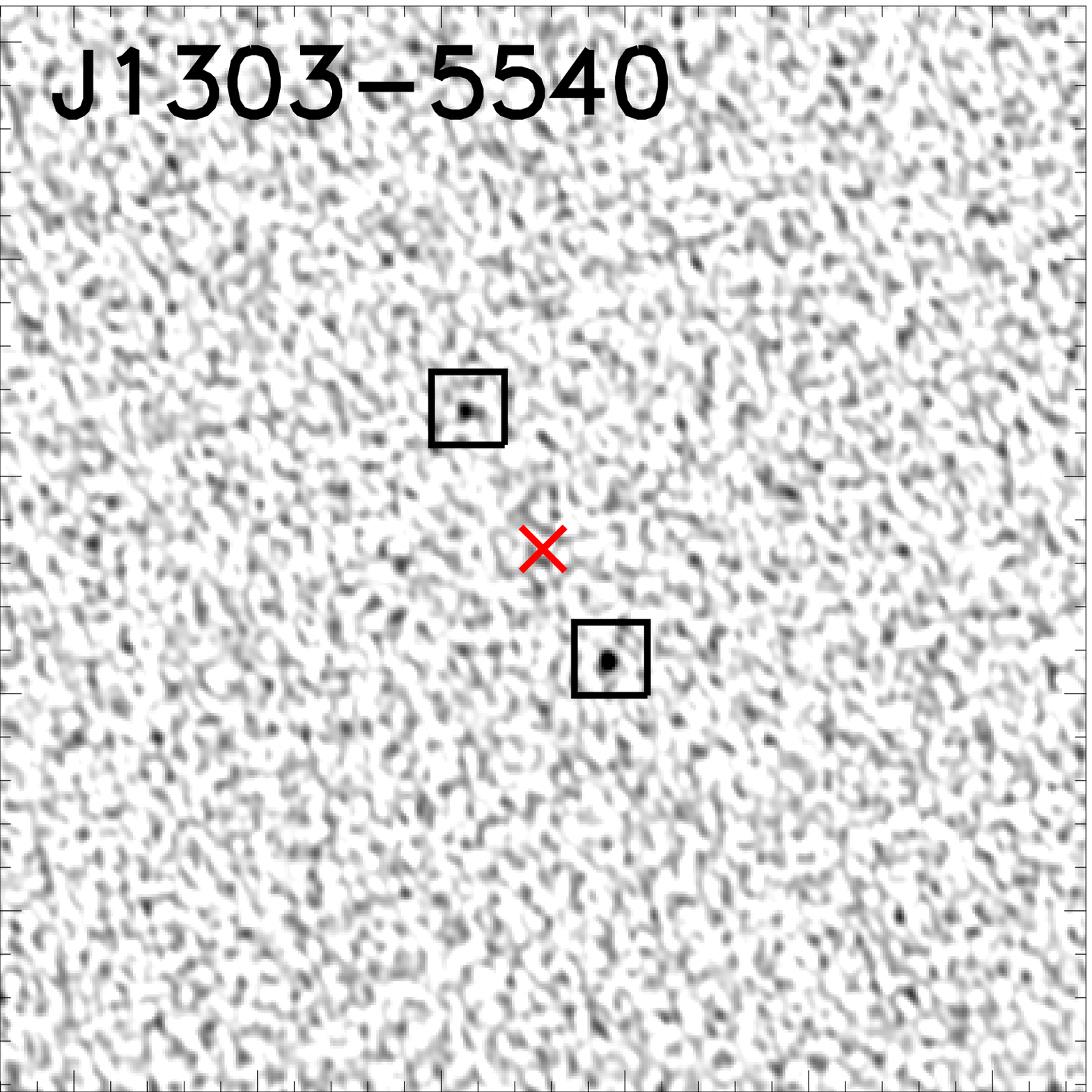} 
\includegraphics[width=0.19\textwidth]{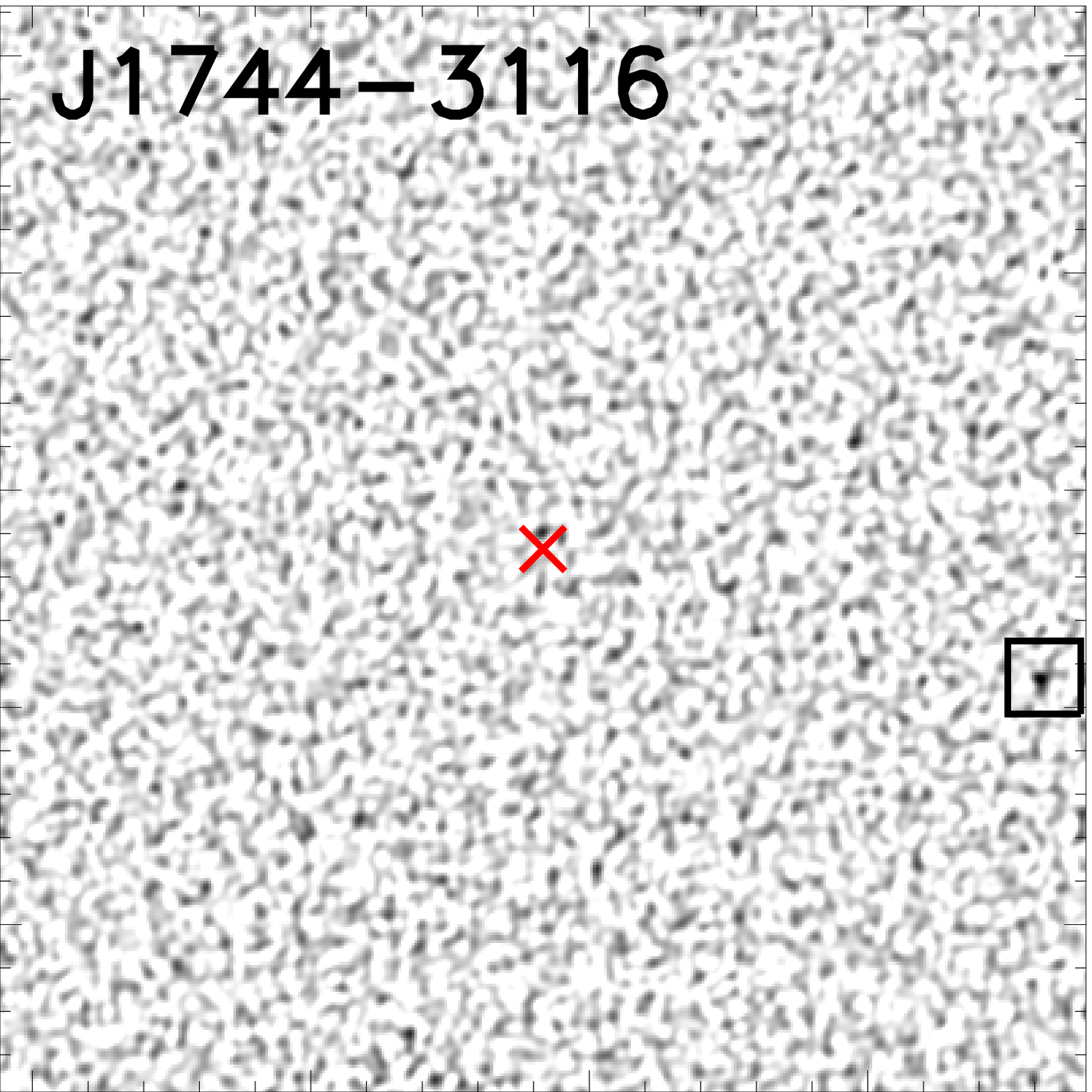} 
\includegraphics[width=0.19\textwidth]{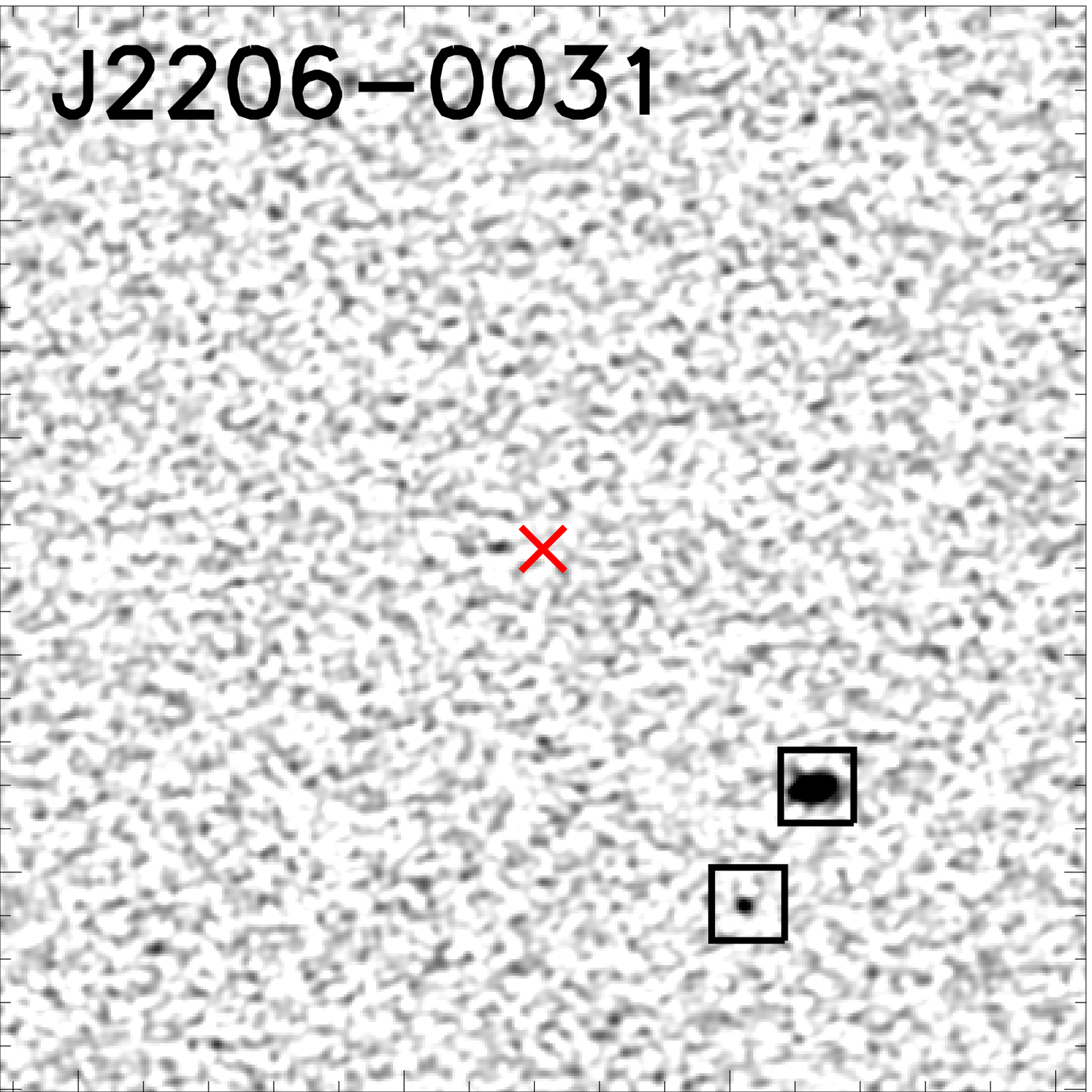} 
\includegraphics[width=0.19\textwidth]{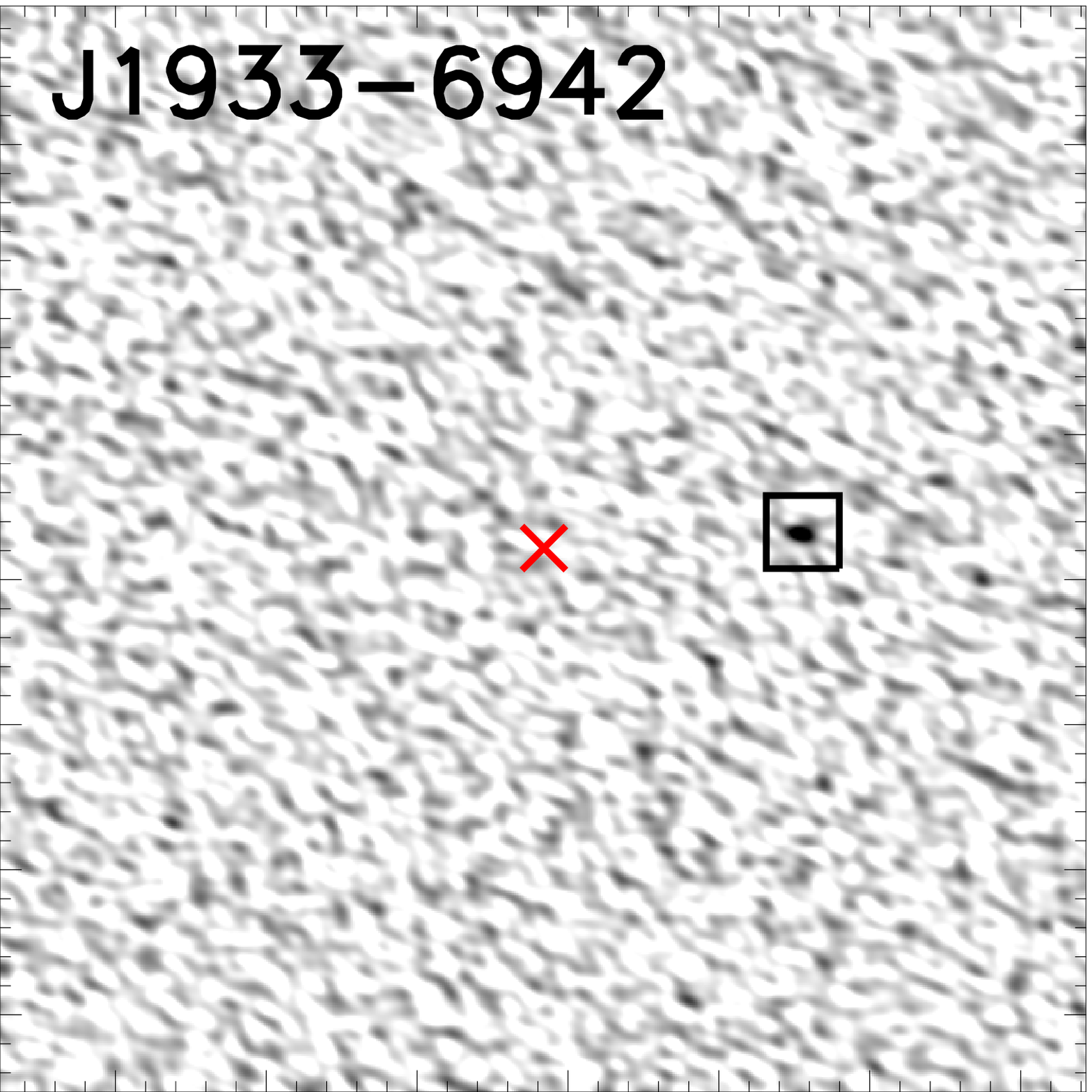} 
\includegraphics[width=0.19\textwidth]{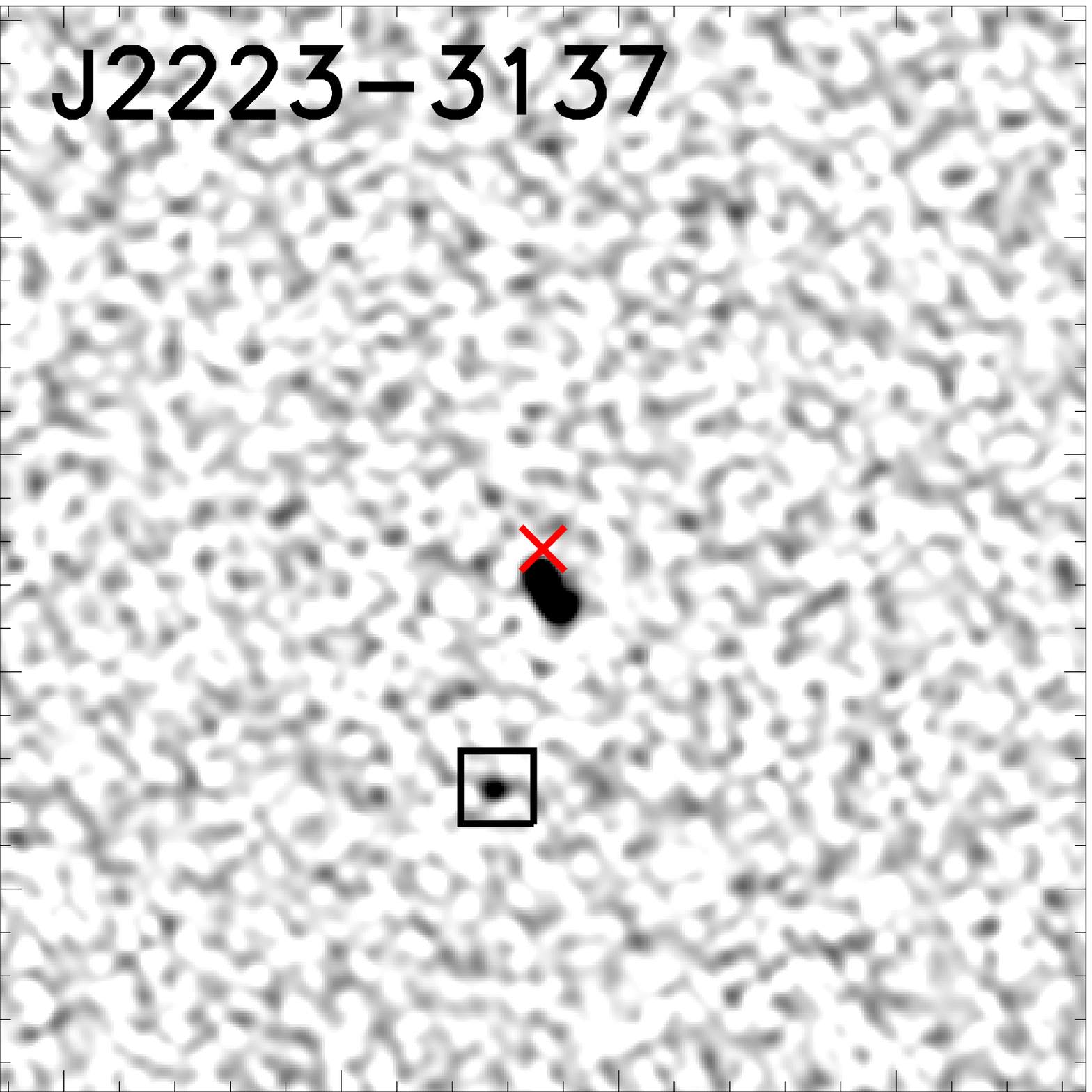} 
\includegraphics[width=0.19\textwidth]{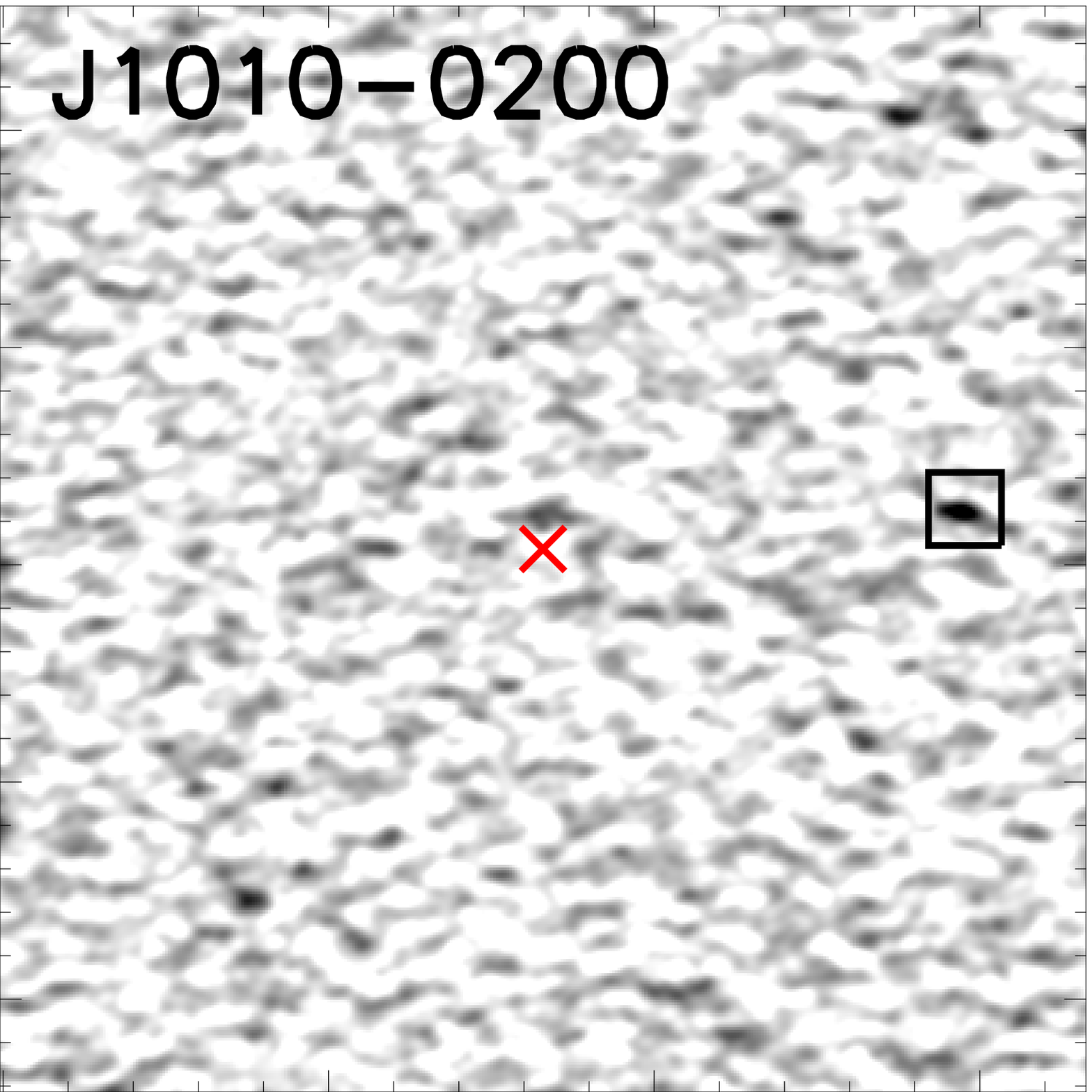} 
\includegraphics[width=0.19\textwidth]{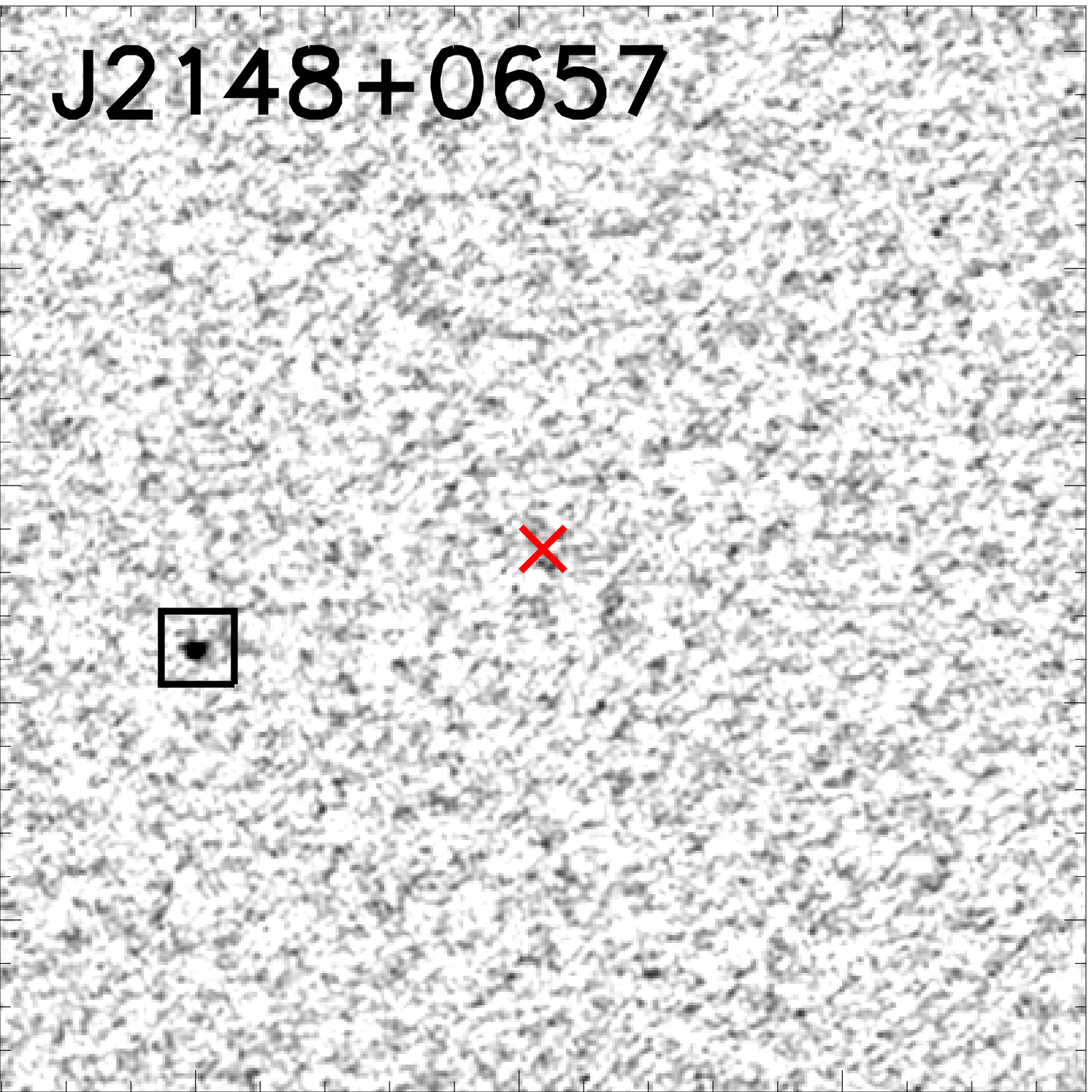} 
\includegraphics[width=0.19\textwidth]{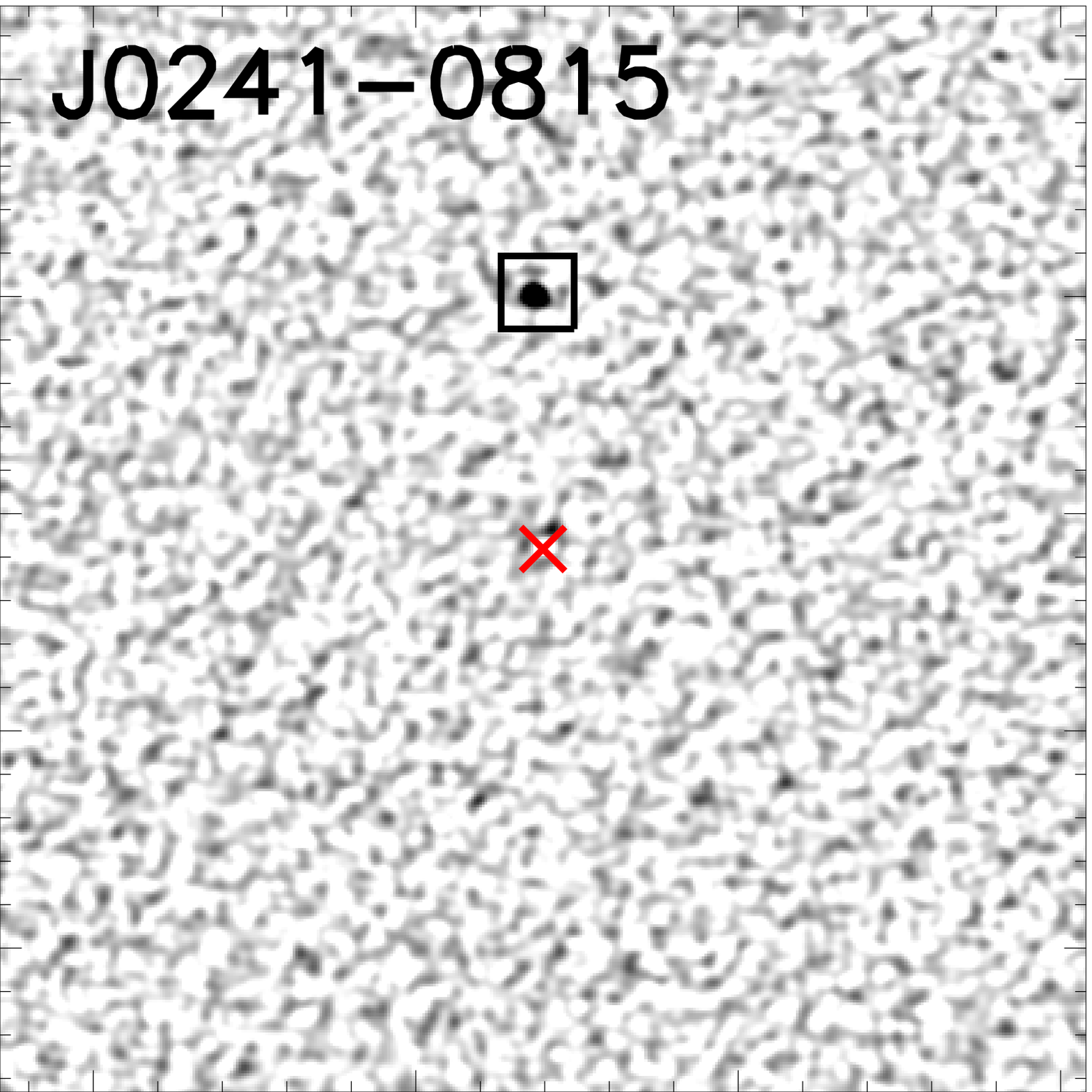} 
\includegraphics[width=0.19\textwidth]{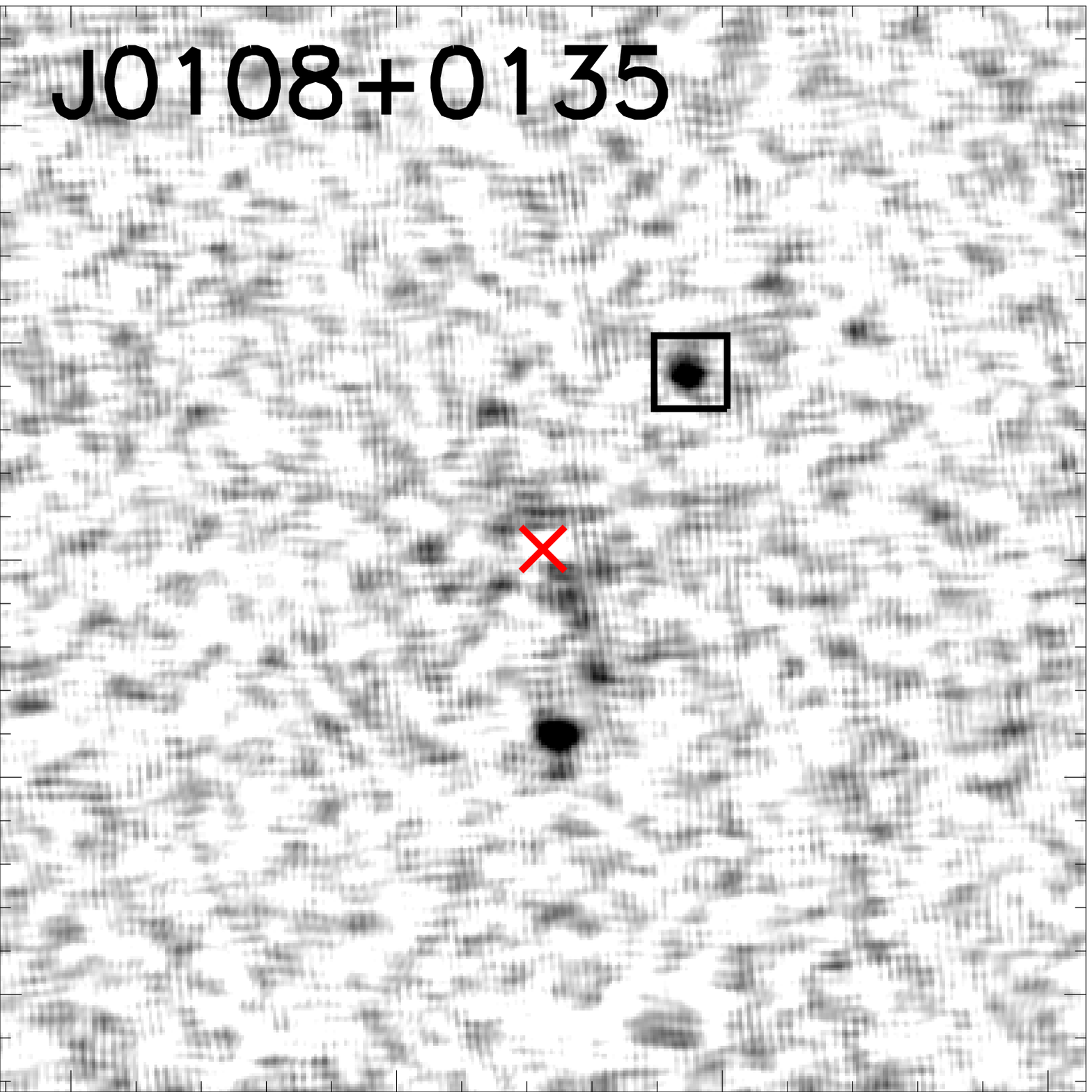} 
\includegraphics[width=0.19\textwidth]{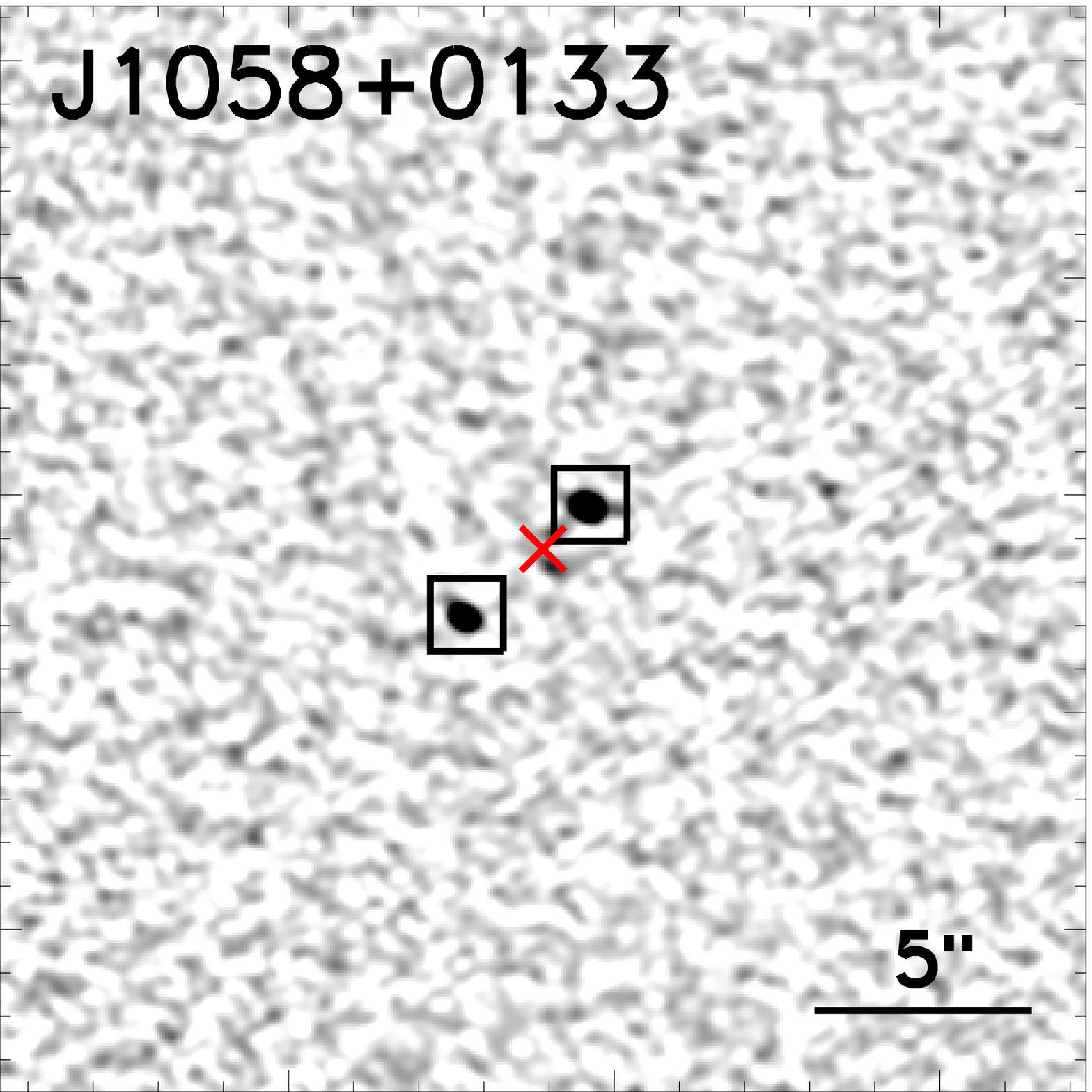} 
\caption{ALMA images of the 13 DSFGs detected around calibrators (represented by the black squares). The calibrators lie at the center of each map, represented by the red cross; they have been subtracted in the $uv$ plane, using point-source models, prior to imaging.  These are 870-$\mu$m (ALMA band-7) images, except in the two cases where only band-6 data are available, shown prior to the correction for PB attenuation. For J0108+0135 we also use the ALMA band-6 image in order to show ALMACAL\,J010838.75+013455.9, part of the jet emanating from the calibrator as revealed by ALMA band-3 imaging.  Each image is 25$''$ on each side ($\sim 1.5 \times$ the {\sc fwhm} of the band-7 PB). The jet emanating from the calibrator, J2223$-$3137, is clearly visible. N is up; E is to the left.}
\label{detected_SMGs_fig}
\end{figure*}

\subsection{Source catalogue}\label{section_source_catalog}

In our 69 ALMA maps we have found eight and 11 submm detections in B6 and B7, respectively. Accounting for all the galaxies detected in B6 and B7, this represents a total sample of 13 submm detections at $> 5 \sigma$. The ten calibrator fields in which our submm emitters were found are shown in Figure~\ref{detected_SMGs_fig}.

To ensure that the flux densities of the detected sources are well determined, we re-imaged the visibilities of the calibrators around which our DSFGs are located using a slightly different procedure.  We defined around each submm detection a cleaning box with 1.5$''$ on each side, a value greater than all our synthesised beams (see Table~\ref{hola_tabla}) and cleaned down to the r.m.s.\ of each dirty map.  As will be explained in \S\ref{section_flux_boosting}, this non-interactive cleaning method provides the most accurate determination of the flux density of the detected DSFGs. After cleaning, the maps were corrected for PB attenuation using {\sc impbcor} and the flux densities and uncertainties were then determined in the PB-corrected maps using {\sc imfit}, with the same box used during cleaning. The coordinates and multi-band flux densities\footnote{Quoted uncertainties include the fitting errors; since we do not assume that sources are point-like, these uncertainties can be larger than the local r.m.s.\ noise in the image.} of the detected DSFGs are presented in Table~\ref{table_detected_real_SMGs}. The $S_{\rm 3.0 mm}$ column shows the ALMA B3 coverage of our sample, where available. None are detected (expressed by the symbol $< \sigma$), meaning their SEDs are all consistent with DSFGs.

\begin{table*}[!t]
\begin{center}
\caption{DSFGs detected up to 2015 July in our ALMA submm survey (see Figure~\ref{detected_SMGs_fig}).}
\label{table_detected_real_SMGs}
\begin{tabular}{l c c c c c c c}
\hline
Source 						&	$S_{\rm 870 \mu m} \, [{\rm mJy}]$	&	$S_{\rm 1.2 mm} \, [{\rm mJy}]$ &	$S_{\rm 3.0 mm} \, [{\rm mJy}]$		  	\\
\hline
ALMACAL\,J130349.05$-$554034.2			&	$0.40 \pm 0.09$	&				 	&								\\
ALMACAL\,J130349.43$-$554028.5			&	$0.66 \pm 0.08$	&				 	&								\\
ALMACAL\,J174422.69$-$311639.4			&	$1.12 \pm 0.27$	&					&  	$< \sigma$					\\
ALMACAL\,J220642.87$-$003108.1			&	$6.89 \pm 0.36$	&	 				&								\\
ALMACAL\,J220642.98$-$003110.8			&	$0.71 \pm 0.15$	&	 				&								\\
ALMACAL\,J193329.46$-$694258.4			&					&	$1.42 \pm 0.20$ 	&								\\
ALMACAL\,J222321.73$-$313707.7			&					&	$0.23 \pm 0.04$ 	&								\\
ALMACAL\,J101051.03$-$020018.8			&	$0.52 \pm 0.15$	&	$0.20 \pm 0.04$ 	&								\\
ALMACAL\,J214806.00+065736.2			&	$2.09 \pm 0.32$	&	$0.55 \pm 0.07$ 	&	$< \sigma$					\\
ALMACAL\,J024104.82$-$081515.0			&	$0.89 \pm 0.10$	&	$0.49 \pm 0.06$ 	&	$< \sigma$					\\
ALMACAL\,J010838.56+013504.2			&	$2.20 \pm 0.15$	&	$0.80 \pm 0.17$ 	&	$< \sigma$					\\
ALMACAL\,J105829.54+013359.7			&	$6.48 \pm 0.30$	&	$2.16 \pm 0.17$ 	&	$< \sigma$					\\
ALMACAL\,J105829.73+013357.2			&	$4.35 \pm 0.18$	&	$1.64 \pm 0.09$ 	&	$< \sigma$					\\
\hline
\end{tabular}
\end{center}
\end{table*}

\begin{figure*}
\centering
\includegraphics[width=0.55\textwidth]{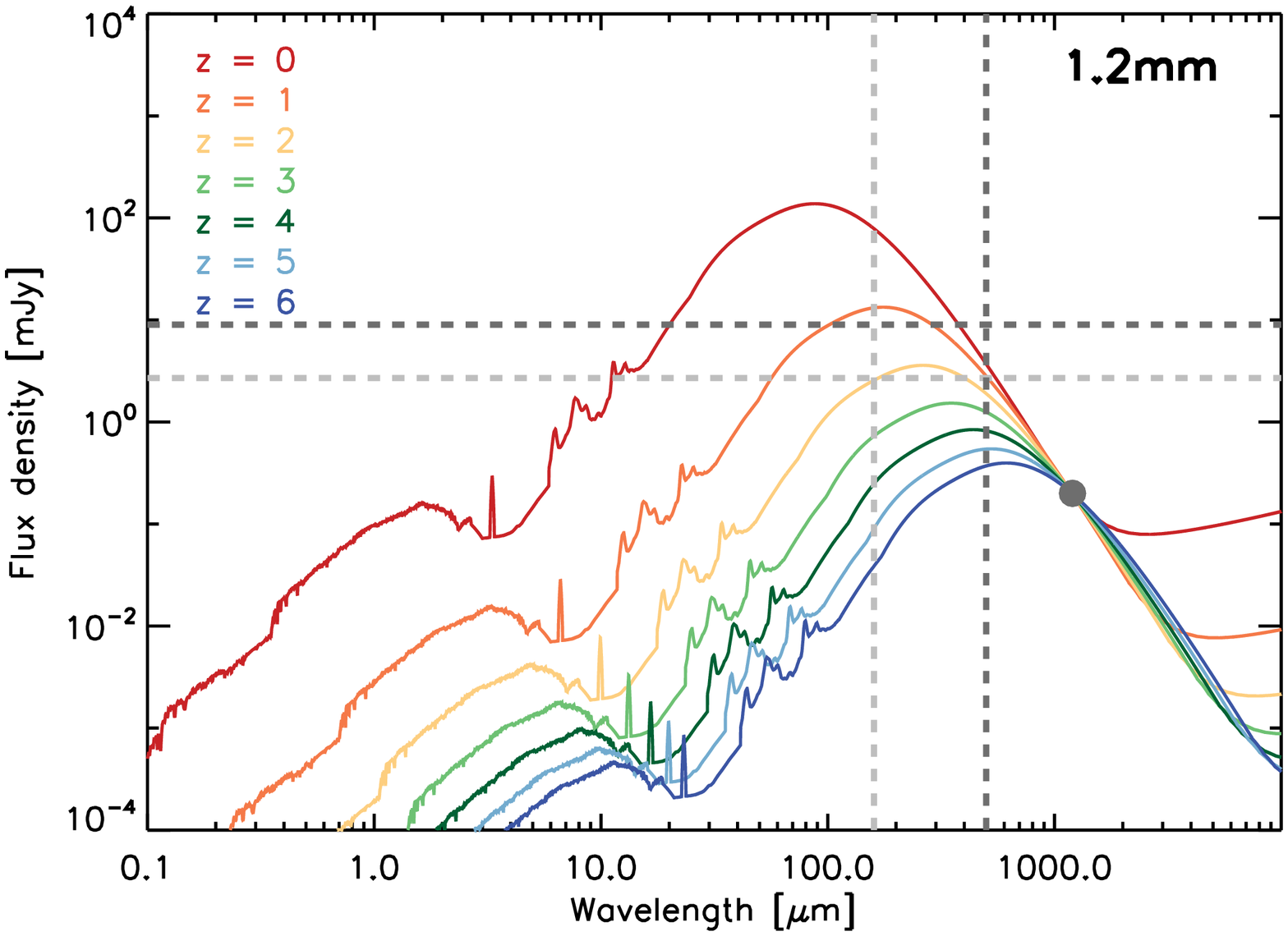} 
\hspace{-20mm}
\includegraphics[width=0.55\textwidth]{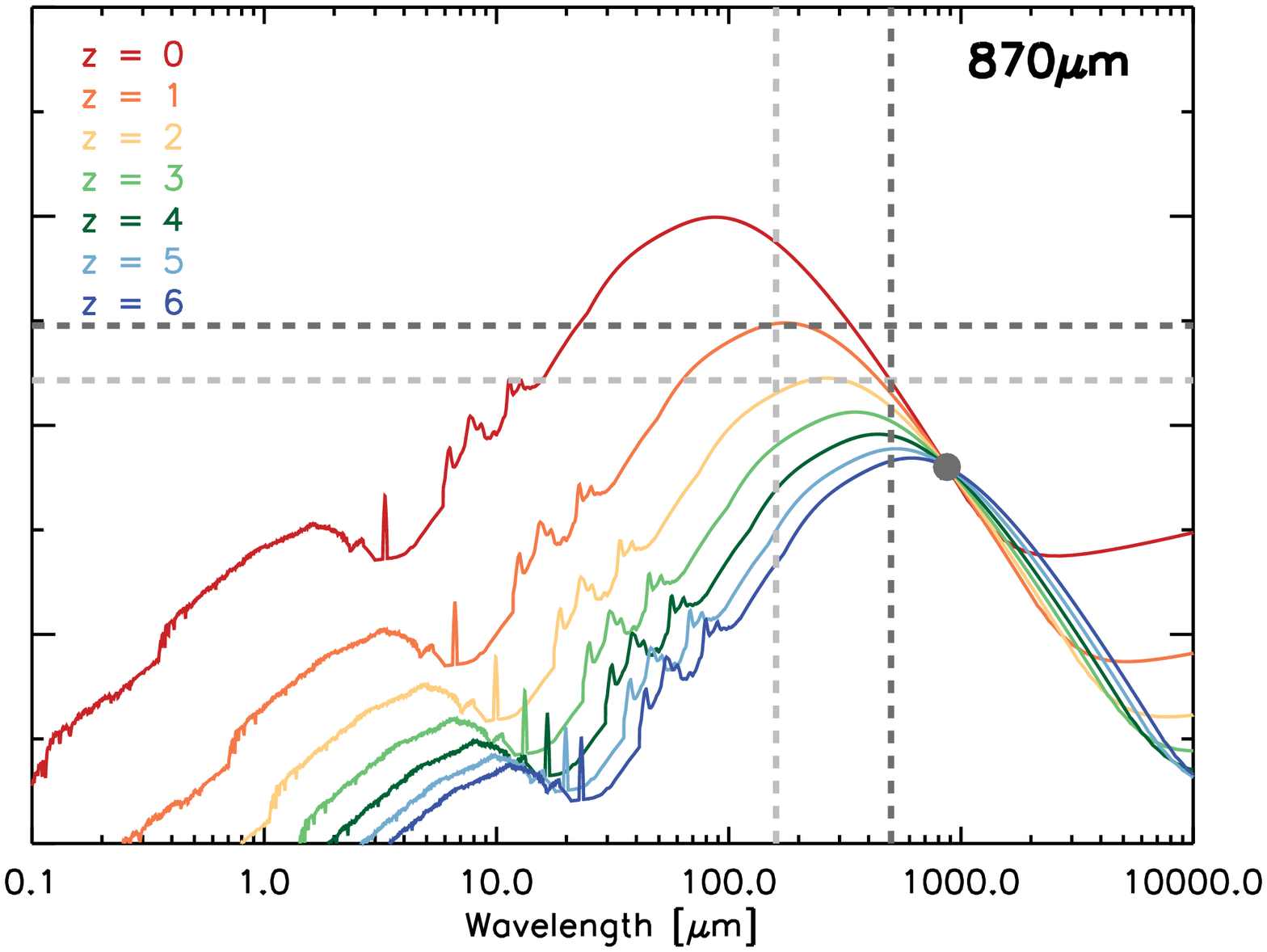} 
\caption{SEDs of our faintest ALMA-selected galaxies -- those detected in B6 (\emph{left}) and B7 (\emph{right}) -- as a function of their possible redshifts. Their observed flux densities are indicated by the grey dots (uncertainties smaller than the size of the dots).  For this plot we have used the median SED for LABOCA sources from the LESS survey \citep{Swinbank2014MNRAS.438.1267S}.  Dashed lines show the limiting PACS 160-$\mu$m and SPIRE 500-$\mu$m flux densities for one of the deepest survey carried out with \emph{Herschel} \citep{Elbaz2011}.  If they lie at $z\geq 2$, the faintest galaxies detected at the present stage of our survey would not have been detected in the deepest observations carried out with \emph{Herschel}.  They likely represent, therefore, a newly discovered population of galaxies, the bridge between \emph{Herschel}- and SCUBA-detected galaxies and the classical UV-selected population.}
\label{FIR_SED_faintest_galaxies}
\end{figure*}

Most of our submm emitters are fainter in B7 (i.e.\ at around 870\,$\mu$m) than typical SCUBA- or LABOCA-detected SMGs.  Ten of our 13 DSFGs are fainter than the faintest deboosted SMG reported in the LESS survey, 3.5\,mJy \citep[][where the detection threshold was ${\rm SNR} \geq 3.75$]{Weiss2009ApJ...707.1201W_LABOCA}. Around half of our DSFGs are fainter than 1\,mJy at 870\,$\mu$m and they are all detected at relatively high SNR and with sub-arcsec spatial resolution.  The median  870-$\mu$m flux density of our DSFGs is comparable with the flux density of Lyman-break galaxies (LBGs) at $z \sim 3$ \citep{Coppin2015MNRAS.446.1293C}, detected only via stacking.  Since we aim to carry out a deep and wide survey, the faintest galaxies detected are of special interest.  Figure~\ref{FIR_SED_faintest_galaxies} shows the FIR SEDs of the faintest galaxies detected in B6 (left panel) and B7 (right panel), respectively.  We do not yet know their redshifts, so we show SEDs for $z = 0$--6, adopting the shape of the median SED of the LESS SMGs \citep{Swinbank2014MNRAS.438.1267S}.  Figure~\ref{FIR_SED_faintest_galaxies} indicates, then, that our faintest DSFGs are fainter than FIR-detected LBGs, FIR-detected $sBzK$ galaxies or FIR-detected H$\alpha$ emitters \citep{Oteo2014MNRAS.439.1337O,Oteo2013A&A...554L...3O,Oteo2015MNRAS.452.2018O} at $z \leq 2$. Indeed, our faintest DSFGs would not have been detected by the deepest survey carried out by \emph{Herschel} and they may thus represent the link between the extreme population of \emph{Herschel} or SCUBA-/LABOCA-selected galaxies and the less extreme UV-selected galaxies. It is clear that our survey represents a significant step towards the discovery and characterisation of faint, sub-mJy DSFGs.

There are also a number of bright DSFGs in our sample, with $S_{\rm 870 \mu m} > 4 {\rm mJy}$. The emission from these galaxies is confined to $\approx$0.3$''$, being only slightly resolved, as suggested by their peak to integrated flux densities. They represent a population of extremely IR-bright galaxies whose prodigious star formation must be confined to a remarkably small volume \citep[see also][]{2015ApJ...799...81S,2014arXiv1411.5038I}.

\begin{figure}
\centering
\includegraphics[width=0.48\textwidth]{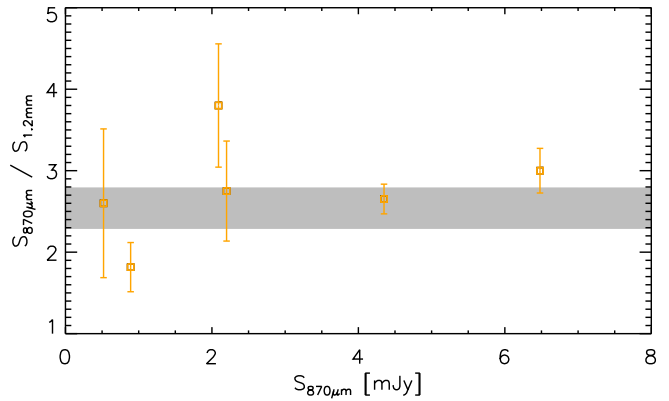} 
\caption{Ratio between flux density at 870\,$\mu$m and 1.2\,mm for the six DSFGs with ALMA detections in both B6 and B7.  The shaded area represents the ratio expected for the composite SED of SMGs from the LESS survey for the redshift range, $z=1$--3 \citep{Swinbank2014MNRAS.438.1267S}. High $S_{\rm 870 \mu m} / S_{\rm 1.2 mm}$ ratios can be explained with a higher dust emissivity index. Low $S_{\rm 870 \mu m} / S_{\rm 1.2 mm}$ ratios could be indicative of higher redshifts.}
\label{spectral_index_SMG}
\end{figure}

Among the 13 detected DSFGs, six have measurements in both B6 and B7.  Figure~\ref{spectral_index_SMG} shows that their $S_{\rm 870 \mu m} / S_{\rm 1.2 mm}$ ratios are generally in good agreement with those expected for DSFGs, as represented by the average SED of SMGs in the LESS survey at redshifts, $z=1$--3 \citep{Swinbank2014MNRAS.438.1267S}. One of the galaxies has an elevated $S_{\rm 870 \mu m} / S_{\rm 1.2 mm}$ which can be explained with an emissivity $\beta \geq 2$, already found for some DSFGs \citep[see for example][]{Casey2011MNRAS.411.2739C}. The lowest $S_{\rm 870 \mu m} / S_{\rm 1.2 mm}$ ratio in our sample is commensurate with that galaxy lying at $z>3$.  This is the first time that FIR/submm spectral indices have been derived for bright or faint DSFGs at matched, high spatial resolution.  The upper limits in $S_{\rm B7,B6} / S_{\rm B3}$  are also compatible with the SEDs of high-redshift DSFGs.

Despite the smaller area covered by our B7 observations (Figure~\ref{area_as_a_funtion_of_flux}) we have detected more galaxies in B7 than in B6.  For the fields where data are available only in a single band, we still detect more galaxies in B7 than in B6.  This apparent discrepancy is due to a combination of different factors, including the different noise levels in the maps, the low-number statistics at the present stage of the survey, the selection function of faint DSFGs as a function of wavelength, and the multiplicity of DSFGs (\S\ref{mullet}).  Among the five DSFGs detected only in B7, four are found in pairs, so the five galaxies are actually detected in only three maps. This is still larger than the number of maps with B6-only DSFGs (two), and merely reflects the need to survey large areas down to low noise levels to derive robust number counts.

\subsection{Multiplicity}
\label{mullet}

Previous studies exploring the environments of bright SMGs have reported that they tend to be strongly clustered \citep{Blain2004ApJ...611..725B,Scott2006MNRAS.370.1057S,Hickox2012MNRAS.421..284H}. Furthermore, it has been reported that bright SMGs found with single-dish telescopes are often resolved into several different components when they are observed at high spatial resolution with ALMA \citep{Karim2013MNRAS.432....2K,Simpson2015ApJ...799...81S}; indeed, \citet{2007MNRAS.380..199I} saw the same effect with the Very Large Array, arguing that many of the pairs must be physically associated rather than chance alignments of dusty galaxies in the same line of sight.

Among our 13 DSFGs, six are found in pairs, with typical separations of a few arcsecs, consistent with these earlier findings. Observed with single-dish telescopes, these pairs would be seen as single, unresolved emission, similar to the classical population of SCUBA- or LABOCA-selected SMGs.  At the present stage of our survey, we can confirm only that the two DSFGs seen around J1058+0133 are at the same redshift (Oteo et al.\ 2015, in preparation).  One of the advantages of using ALMA calibrators to study the DSFG population is that the calibrators will eventually be observed in a frequency range where there is one or more emission lines. A single line will allow us to confirm whether the multiple components lie at the same redshift \citep[e.g.][]{Chapman2015MNRAS.449L..68C}.

\subsection{Caveats}

As we have described, there are a number of advantages relating to the use of ALMA calibration data to study DSFGs.  Nevertheless, we must note  some caveats.

The first relates to the possibility that unresolved, high-redshift DSFGs may be confused with jets emanating from the calibrators.  ALMA calibrators are typically blazars with strong jets that are currently oriented along our line of sight. Indeed, jet signatures are clearly visible in some of our ALMA maps.  Jets tend to present an extended, cometary shape and sometimes curved tails emanating from the calibrator (see, for example, J2223$-$3137 in Figure~\ref{detected_SMGs_fig}). Due to their morphology, these jets are relatively easy to identify and thus distinguish from high-redshift DSFGs by visual inspection of the clean maps, since their extended emission points towards the calibrator.  However, there may also be cases where the jet coming from a calibrator may appear unresolved. This unresolved emission could be confused with a high-redshift DSFG. It is, however, possible to discriminate between jets and DSFGs using the spectral index of the emission in the ALMA data.  The ratio between ALMA B6 or B7 and B3 or B4 flux densities provide useful diagnostics to distinguish between thermal and non-thermal emission: $S_{\rm B3,4} > S_{\rm B6,7}$ and $S_{\rm B7} < S_{\rm B6}$ would be typical for a jet.  When there are no ALMA B3 or B4 data, the flux density ratio between B6 and B7 can also be used to identify jets, though DSFGs at very high redshift, or very cold galaxies, may also have $S_{\rm B7} < S_{\rm B6}$ due to the dust emission peak shifting to redder wavelengths; we have not applied such a cut.  We have, however, discarded from our sample any submm detections with $S_{\rm B3,B4} > S_{\rm B6,B7}$.  A clear example of the possible confusion between jets and DSFGs is ALMACAL\,J010838.75+013455.9, one of two submm detections around the calibrator, J0108+0135.  These two submm detections were first identified using ALMA B6 data and they both resembled DSFGs. However, ALMA multi-band data revealed that ALMACAL\,J010838.75+013455.9 is not detected in B7, contrary to the expectations for the SED of a high-redshift DSFG, despite the fact that B7 noise level would have allowed a detection.  It also exhibits $S_{\rm 3.0 mm} > S_{\rm 1.2 mm}$, indicating that it is likely related to jet emission from the calibrator.  It is therefore excluded from our final sample of DSFGs in Table~\ref{table_detected_real_SMGs}.  On the other hand, ALMACAL\,J010838.56+013504.2 is confirmed to be a distant DSFG due to its B7/B6 flux density ratio and its non-detection in B3.  Two other clear examples of contamination lie around J1733$-$1304 and J0522$-$3627; despite lying far from the calibrators, high $S_{\rm B3,4} / S_{\rm B6,7}$ ratios are clearly indicative of non-thermal emission.

Turning now to a second caveat, over-densities of SMGs have been found around bright radio galaxies and quasars on a variety of scales \citep[e.g.][]{2000ApJ...542...27I,2008MNRAS.390.1117I,2003Natur.425..264S, 2004ApJ...604L..17S}.  It is reasonable, then, to worry that our ALMA calibrators might also be related to DSFG over-densities.  In defense of our work, the properties of our calibrators -- in particular their mass -- are not as extreme as radio galaxies \citep[see][]{2007ApJS..171..353S}, which are the most massive galaxies at any redshift, and which in any event are associated with over-densities which have proved difficult to prove conclusively; our calibrators are blazars, which are bright because we are looking down the throat of their jets, not because they are hosted by spectacularly massive galaxies. Nevertheless,  we concede that selection of DSFGs around bright submm calibrators may be biased at a low level.  Indeed, a stronger bias is present in most if not all previous reports of faint ALMA counts in the literature, since those previous works often derive number counts using ALMA science data that were deliberately centered on IR-bright or otherwise extreme galaxies that might also be associated with over-dense regions.

An entirely unbiased determination of submm number counts (see \S\ref{section_number_counts}) will only be possible when we can image large areas of the sky. While deep maps are possible with ALMA, wide observations are time-consuming due to the small FoV of the ALMA antennas in the bands that probe dust emission from star-forming galaxies.

\subsection{Spurious sources}\label{subsection_comments_on_spurious_sources}

Creating large samples using a low SNR threshold, ${\rm SNR \sim 3-4}$, leads inevitably to the inclusion of spurious sources.  Working with samples contaminated by spurious sources can be a valid approach if our interest lies only in the study of number counts, since we can determine and apply accurate correction factors.

Here, instead, we have opted to be conservative -- to reduce as much as possible the contamination from spurious sources -- since one of our aims is to study the nature of the detected DSFGs.  For this we need to be sure that the sources we detect are real, lest we were to identify a spurious source as a dropout in another band, perhaps initiating a long and sorry saga.  In our source selection, therefore, we have selected sources at ${\rm SNR} \geq 5$, a threshold higher than has been adopted in all recent reports on faint ALMA counts \citep{Ono2014ApJ...795....5O,Fujimoto2015arXiv150503523F,Carniani2015arXiv150200640C,Hatsukade2013ApJ...769L..27H}.  While this will significantly decrease the contamination from false detections in our final sample of DSFGs, we have inevitably excluded some of the faintest DSFGs from our catalogue.  While we can be reasonably confident that the number of false detections in our resulting $5\sigma$ sample should be zero, we can state with absolute confidence that all six of the DSFGs with multi-band detections, as discussed above and shown in Table~\ref{table_detected_real_SMGs}, are real.


\subsection{Completeness}\label{section_completeness_inject_galaxies}

We have employed the traditional method of injecting simulated sources with different flux densities at random positions of our ALMA maps in order to study the completeness of our selection method as a function of SNR.  All previous works in this area have added simulated sources to the clean maps.  Since we are working with interferometric data we have instead opted to perform our simulations in the visibility plane.  We have taken ALMA calibrator maps where no DSFG is detected and inserted artificial, point-like sources with flux densities ranging from 0.1 to 10\,mJy in steps of 0.025\,mJy, thus covering a wide range of SNRs. Sources were always injected within $1.5\times$ {\sc fwhm}$_{\rm PB}$, the same area used for source detection (\S\ref{section_source_selection}). We repeated the source injection 100 times for each value of the input flux density.  Once a source has been injected at a given position, we shifted the phase center of the map to those coordinates to ensure the source should show up at the center of the dirty map.  Because this is done for all the injected sources, it is possible to define exactly the same cleaning box for all of them, ensuring that the cleaning, source detection, and flux-density determination is a homogenous and consistent process.  Imaging and cleaning was accomplished in non-interactive mode, down to the r.m.s.\ of the dirty map.  As will be shown in \S\ref{section_flux_boosting}, this cleaning method provides accurate values of the recovered flux density of the simulated sources and also represents the optimal method for measuring the flux density of our detected DSFGs.  We consider than an injected source has been recovered if it is detected with SExtractor at $\geq 5\sigma$ within a synthesised beam of the center of each map.  Clean maps, prior to PB correction, are used for source detection, consistent with the process used to select our DSFGs (see \S\ref{section_source_selection}).

As a result of our simulations we have determined that our survey is nearly 100\% complete at ${\rm SNR} \geq 7$ and $\sim$80\% complete at ${\rm SNR} \geq 6$. These SNR values are higher than those reported in most previous works that have analysed faint DSFGs detected in deep ALMA maps.  For example, the completeness in \cite{Simpson2015ApJ...807..128S} is 93\% at $> 4 \sigma$, rising to about 100\% at $5.5 \sigma$, meanwhile \citet{Fujimoto2015arXiv150503523F} claim their ALMA observations are 90\% complete at ${\rm SNR} > 4.5$ (their Figure~3).  The higher completeness at a given SNR derived in previous works is a consequence of the lower SNR threshold used for source selection.  Previous works have employed selection thresholds at $\rm SNR \geq 4$, or lower, which increases the completeness at low flux densities (or SNRs) at the cost of increasing the number of spurious detections.

\subsection{Effective area}\label{section_area_covered_by_the_survey}

The sensitivity of any single-pointing ALMA image (or any interferometric image) decreases with increasing distance from the center of the map due to the PB response of the individual antennas. The effective area sensitive to a given flux density therefore varies with flux density. As an example, a galaxy detected in the center of the map at $6\sigma$ would not be detected at $\geq 5\sigma$ were it located instead at the edge of the map (defined by $1.5 \times$ {\sc fwhm}) and would not then be included in our sample. 

\begin{figure}
\centering
\includegraphics[width=0.48\textwidth]{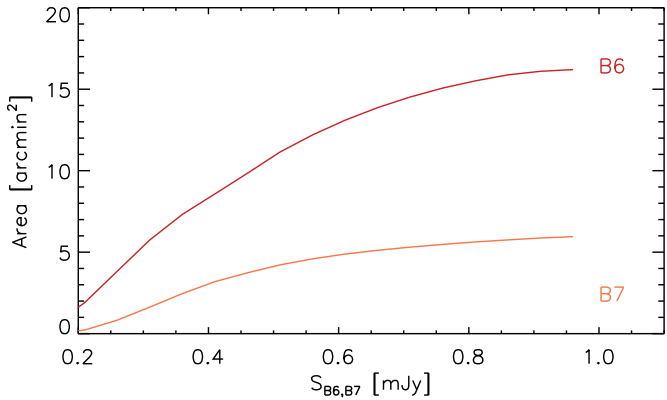} 
\caption{Effective area covered by our ALMA submm survey as a function of the flux density of the detected sources for the two bands considered in this work.  The sensitivity of a given ALMA calibrator map is a function of the distance to the center of the map.  The effective area covered by the survey is then a function of the flux density of the detected sources.  For a given source detected at a given band with a given flux density, we calculate the area where that galaxy could have been detected at $\geq 5\sigma$ (our limit for source detection -- see \S\ref{section_source_selection}) in all pointings in that band.  In this calculation we have modelled the PB response (given by {\sc casa} during imaging) with a Gaussian.}
\label{area_as_a_funtion_of_flux}
\end{figure}

Obtaining the relationship between the effective area of our survey and the flux density of the sources detected is important since we have to correct the number counts for this effect (see \S\ref{section_number_counts}). For a given flux density, $S_{\rm in}$, and a map with an r.m.s.\ of $\sigma_{\rm in}$ we calculated the SNR at the center of the map, ${\rm SNR_{\rm cen} = S_{\rm in} / \sigma_{\rm in}}$. We then obtained the radius, $r_{\rm lim}$, at which that SNR decreases to five, which gives the area in which a galaxy with a flux density, $S_{\rm in}$, can be detected.  If $2 \times r_{\rm lim}$ exceeds $1.5 \times$ {\sc  fwhm}, we define  $r_{\rm lim}$ from $2 \times r_{\rm lim} = 1.5 \times$ {\sc fwhm}. We performed this calculation for all observations in B6 and B7 independently, since we will report number counts for both bands. Figure~\ref{area_as_a_funtion_of_flux} shows the effective area of our survey as a function of the flux density of the detected sources in B6 and B7.  The total number of pointings in B7 is smaller than in B6; taken together with the smaller area of each B7 observation, this means that the total area covered in B6 (around $16 \, {\rm arcmin^2}$) is considerably higher than that covered in B7 (around $6 \, {\rm arcmin^2}$).

\subsection{Flux boosting}\label{section_flux_boosting}

It has long been known that the flux densities of galaxies detected at relatively low SNR can be boosted due to the presence of noise fluctuations.  We have analysed the effect of flux boosting for our ALMA detections using the same set of simulated point-like sources used in \S\ref{section_completeness_inject_galaxies}. Once the phase center of the map has been shifted to the position of each injected source, and the visibilities have been imaged, as explained in \S\ref{section_completeness_inject_galaxies}, the flux densities of the detected galaxies are measured in the clean maps using {\sc imfit} prior to correction for PB attenuation. 

\begin{figure*}
\centering
\includegraphics[width=0.55\textwidth]{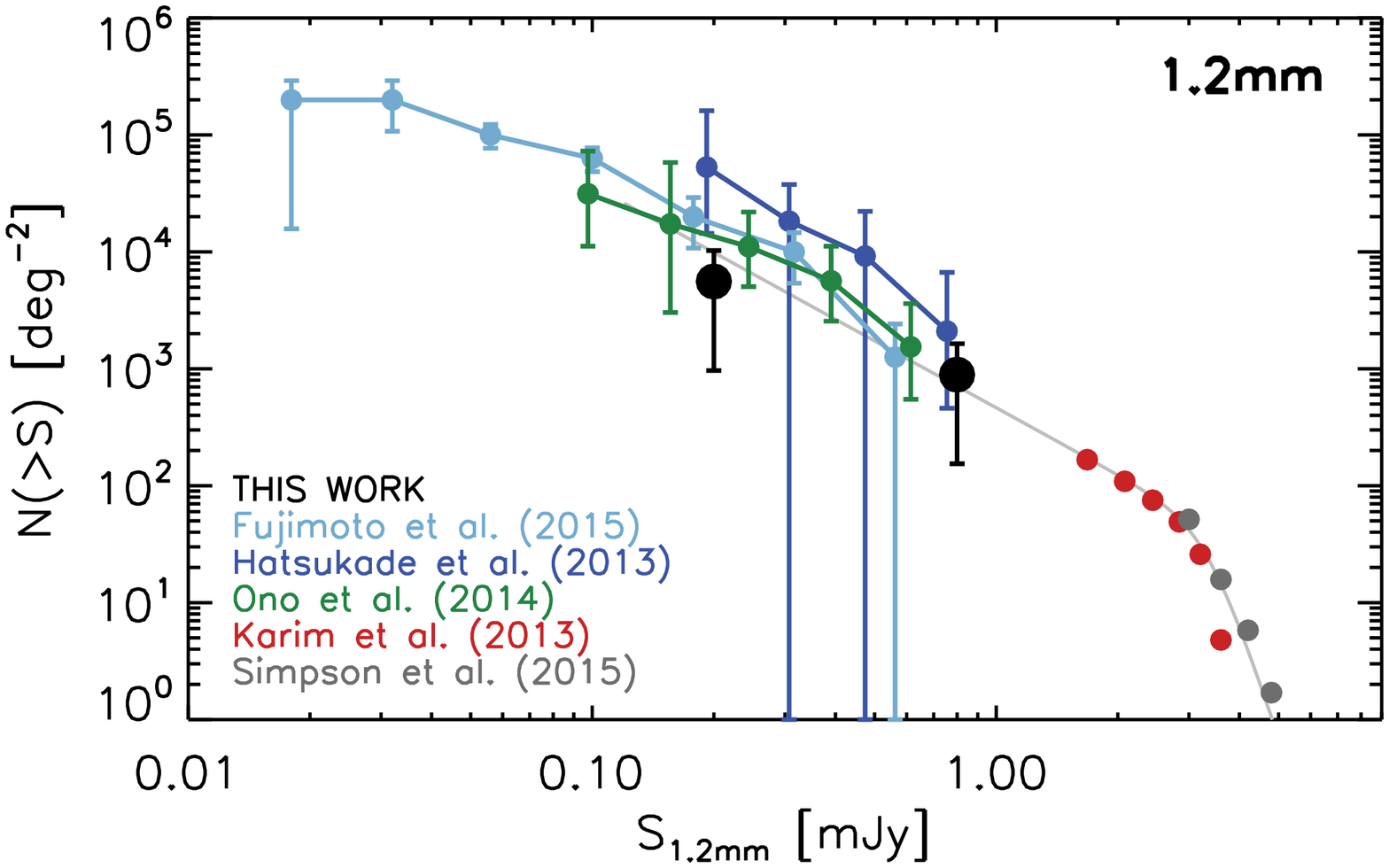}
\hspace{-21mm}
\includegraphics[width=0.55\textwidth]{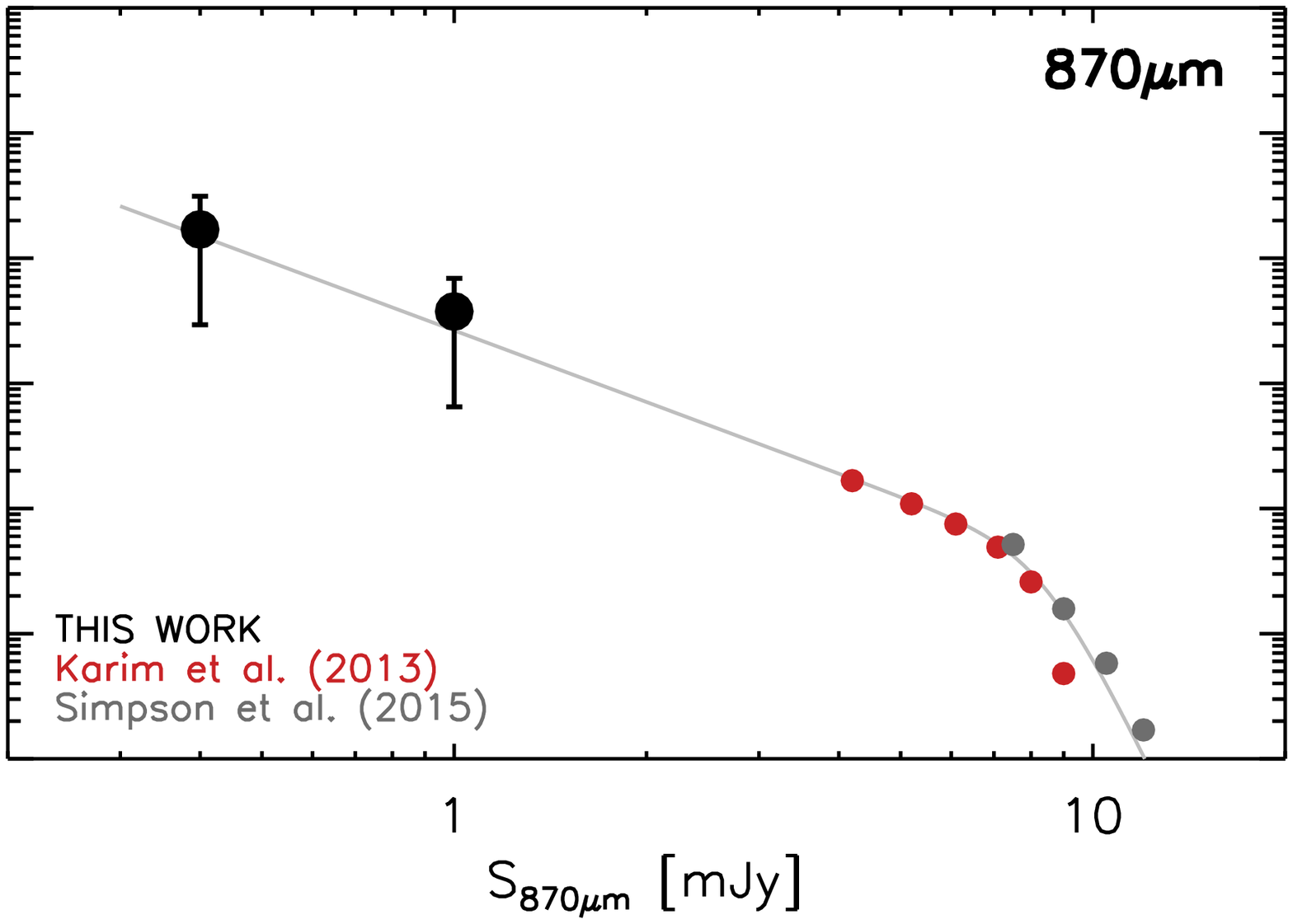}
\caption{Cumulative number counts of DSFGs derived in this work at $870 \, {\rm \mu m}$ (\emph{left}) and 1.2\,mm (\emph{right}). We also plot the results reported by \citet{Hatsukade2013ApJ...769L..27H,Karim2013MNRAS.432....2K,Ono2014ApJ...795....5O,Fujimoto2015arXiv150503523F,Simpson2015ApJ...807..128S}. Number counts determined in previous work at different wavelengths have been converted into 870-$\mu$m and 1.2-mm counts assuming the typical SED of bright SMGs at $z = 2.3$ \citep{Swinbank2014MNRAS.438.1267S}. The grey curve is the fit obtained in \citet{Simpson2015ApJ...807..128S} extrapolated towards the flux densities covered in our work.  Our counts are lower than those presented in previous works; specifically, our counts are lower by a factor of around $2\times$ with respect to \cite{Fujimoto2015arXiv150503523F} and by a factor of about $7\times$ with respect to \cite{Hatsukade2013ApJ...769L..27H}.}
\vspace{9mm}
\label{fig_number_counts}
\end{figure*}

We find that the flux density is over-estimated by up to 20\% for sources detected at ${\rm SNR} \sim 5$, which is in good agreement with the findings of \cite{Simpson2015ApJ...807..128S}.  By contrast, our results are in marked disagreement with the findings of \citet{Fujimoto2015arXiv150503523F}, where $S_{\rm out} / S_{\rm in}$ ratios are reported to be almost unity with a variance of only 5\%.  They argue that their low contamination from flux boosting is due to the high spatial resolution of their ALMA images.  However, we reach a similar spatial resolution as \citet{Fujimoto2015arXiv150503523F} and find that the effect of flux boosting is significant at low SNR. Additionally, \cite{Fujimoto2015arXiv150503523F} considered detections down to ${\rm SNR} \sim 3.7$, where the effect of the flux boosting should be $\gtrsim$30\% if we extrapolate from our $\geq 5 \sigma$ results to lower SNR. Many of our ALMA-detected DSFGs were detected in the SNR regime where the effect of flux boosting is not significant (see Table~\ref{table_detected_real_SMGs}).  The flux densities of our DSFGs detected at ${\rm SNR} < 10$ have been corrected for the boosting effect.


The analysis of the ratio between input and output flux densities of the injected sources also shows that $S_{\rm out} / S_{\rm in} \sim 1$ for sources detected at $>10\sigma$, confirming that our non-interactive imaging method provides robust flux densities.  On the other hand, we obtain that the variation in $S_{\rm out} / S_{\rm in}$ from galaxy to galaxy at low SNR can be significant (up to a factor of two of difference), meaning that caution should be exercised when interpreting individual flux density ratios.


\section{Number counts}\label{section_number_counts}

\begin{table}
\begin{center}
\caption{Cumulative number counts derived in our work.}\label{table_counts_cumul_our_work}
\begin{tabular}{c c}
\hline
$S_{\rm 870 \mu m} \, {\rm [mJy]}$	&	$ N (> S_{\rm 870 \mu m}) [\times 10^3\, {\rm deg^{-2}}]$	\\
\hline
0.4 & $17.0^{14.3}_{14.0}$ \\
1.0 & $3.8^{3.2}_{3.1}$ \\

\hline
$S_{\rm 1.2 mm} \, {\rm [mJy]}$	&	$ N (> S_{\rm 1.2 mm}) [\times 10^3\, {\rm deg^{-2}}]$	\\
\hline
0.2 & $5.6^{4.7}_{4.6}$ \\
0.8 & $0.9^{0.8}_{0.7}$ \\

\hline
\end{tabular}
\end{center}
\end{table}

In this section we present the cumulative number counts obtained for our survey of DSFGs and compare them to previous results published in the literature. Since we have a reasonable number of submm detections in B6 and B7, we present for the first time ALMA cumulative number counts for each band independently. 

The contribution to the cumulative number counts of a source, $i$, with a flux density, $S_i$, is:

\begin{equation}
	N_i(S_i) = \frac{1-f_{\rm sp(S_i)}}{C(S_i) \cdot A(S_i)}
\end{equation}

\noindent
where $f_{\rm sp} (S_i)$ is the fraction of spurious sources at $S_i$, $C(S_i)$ is the completeness of the survey at $S_i$, and $A(S_i)$ is the effective area covered by our survey at $S_i$.  The completeness and effective areas at different flux densities are taken from \S\ref{section_completeness_inject_galaxies} and \S\ref{section_area_covered_by_the_survey}, respectively.  Thanks to our multi-band and multi-epoch observations and our conservative selection criterion, the fraction of spurious sources is expected to be zero: $f_{\rm sp} = 0$ for any $S_i$.  In order to calculate the cumulative number counts at different flux densities, we have to sum over all the galaxies with flux densities higher than the adopted values:

\begin{equation}
	N(>S) = \sum_{\forall S_i > S}\frac{1-f_{\rm sp}(S_i)}{C(S_i) \cdot A(S_i)}
\end{equation}     

Figure~\ref{fig_number_counts} shows the cumulative number counts derived in this work at 1.2\,mm and 870\,$\mu$m, where the flux boosting, completeness, and effective area factors have been included (see also Table~\ref{table_counts_cumul_our_work}). For each band, we have reported number counts at flux densities that evenly divide the sample into two bins with similar numbers of galaxies.  Previous results based on ALMA data, from \cite{Karim2013MNRAS.432....2K}, \cite{Simpson2015ApJ...807..128S}, \cite{Hatsukade2013ApJ...769L..27H}, \cite{Ono2014ApJ...795....5O} and \cite{Fujimoto2015arXiv150503523F} are also shown.  We have decided not to compare with cumulative number counts derived with ground-based single-dish telescopes or {\it Herschel} since bright SMGs are sometimes resolved into multiple components, which may strongly affect the derivations of number counts \citep{Karim2013MNRAS.432....2K,Simpson2015ApJ...807..128S}. 

The number counts derived in previous work have been determined in different ALMA bands.  Cumulative number counts derived from ALMA follow-up observations of SCUBA- or LABOCA-selected SMGs have been determined in B7.  Meanwhile, previous work reporting faint ALMA counts using archival ALMA science data have largely been focused on B6 data due to the larger instantaneous FoV.  In order to avoid uncertainties due to converting flux densities between different bands, we have compared only results obtained in the same ALMA bands, with one exception.  In order to include bright counts in B6, we also include the results from \cite{Simpson2015ApJ...807..128S} and \cite{Karim2013MNRAS.432....2K}.  The counts have been converted from 870\,$\mu$m to 1.2\,mm using the composite SMG SED from the LESS survey \citep{Swinbank2014MNRAS.438.1267S}, redshifted to $z = 2.3$, the median redshift of LABOCA-detected SMGs \citep{Simpson2014ApJ...788..125S}.  This conversion factor is the same as that used in \cite{Simpson2015ApJ...807..128S} and similar to that used in \cite{Fujimoto2015arXiv150503523F}.  We have similarly converted the 1.3-mm number in \cite{Hatsukade2013ApJ...769L..27H} to 1.2\,mm.

Any comparison of number counts between different works will, of course, be affected by this transformation. The SED in \cite{Swinbank2014MNRAS.438.1267S} was derived for bright SMGs, whereas many of our DSFGs are faint and might thus have different dust temperature distributions. Also, the median redshift of the faint DSFGs in our sample could differ from the bright sample of \cite{Simpson2014ApJ...788..125S}, and the redshift distribution also likely depends on the wavelength where the DSFGs were selected \citep{Vieira2013Natur.495..344V}.

It can be seen from the left panel of Figure~\ref{fig_number_counts} that the cumulative 1.2-mm number counts we obtain at the present stage of the survey are lower than previous claims. Even for the case with the best agreement, \cite{Fujimoto2015arXiv150503523F}, we obtain lower number counts by a factor of $2\times$.  A more severe disagreement is found with \cite{Hatsukade2013ApJ...769L..27H}, who report number counts significantly higher than ours.

One of the main differences between our work and previous work is the source-detection procedure.  We selected only galaxies detected at $\geq 5\sigma$, while all previous work reporting faint number counts have included sources detected at lower SNR.  This might lead to the inclusion of spurious detections, as each of these previous papers pointed out, whereas our process leads to negligable contamination.  A fair comparison with published number counts would require consideration of only the $> 5\sigma$ detections in those works.  However, most do not give the flux density, SNR, area covered, and completeness for individual detections. Furthermore, comparison with previous work requires a re-calculation of the area covered and completeness of previous studies, since these depend upon the source detection criteria employed. \cite{Ono2014ApJ...795....5O} did report the list of detections, including their SNR.  Among their 11 submm detections in ten ALMA maps, only two are detected at $>5\sigma$, only one of them as bright as the DSFGs in our sample.

The right panel of Figure~\ref{fig_number_counts} represents the cumulative number counts derived in this work at 870 $\mu$m.  Since all previous work studying faint number counts with ALMA has been focused on lower frequencies to take advantage of the larger instantaneous FoV, our results represent the first determination of faint number counts at 870\,$\mu$m with ALMA.  In Figure~\ref{fig_number_counts} we combine our counts with those derived in \cite{Simpson2015ApJ...807..128S} and \cite{Karim2013MNRAS.432....2K}.  \cite{Simpson2015ApJ...807..128S} fitted a double power law to their own data, and those of \cite{Karim2013MNRAS.432....2K}. It can be seen that the extrapolation of their best-fit function toward $S_{\rm 870 \mu m} \sim 0.3 \, {\rm mJy}$ reproduces the cumulative number counts obtained in this work.

Our wide and deep submm survey, exploiting ALMA calibrator data, is proving to be very powerful for deriving faint number counts, both in ALMA B6 and B7.  This is thanks to the low noise levels that can be reached when combining data for a specific calibrator and the area that can be covered by combining deep data for different calibrators.  As more data become available for a larger number of calibrators, we expect to derive more robust number counts in B6 and B7, based on larger samples, obtained with deeper observations over larger areas.  The large number of calibrators used during ALMA observations will enable us to determine robust number counts even in B8, despite the relatively small FoV.  The combination of number counts in all these bands will provide strong constraints on models of galaxy formation and evolution.


\section{Conclusions}\label{concluuuuu}

In this paper we have presented a novel technique which exploits publicly available ALMA calibration data to carry out a deep and wide submm survey, sensitive to dusty, star-forming galaxies.   Our main conclusions are:

\begin{enumerate}
	\item We have demonstrated the power of using ALMA calibrators to detect and study high-redshift DSFGs by analysing relatively deep data, down to ${\rm r.m.s. \sim 25 \, {\rm \mu Jy \, beam^{-1}}}$, of 69 calibrators. We have focused on ALMA bands 6 ($\sim 1.2 \, {\rm mm}$) and 7 ($\sim 870 \, {\rm \mu m}$), since they probe the dust emission of high-redshift star-forming galaxies. We have covered about 16 and 6\,arcmin$^2$ in ALMA B6 and B7, respectively.
		
	\item Using a conservative detection threshold, $\geq 5\sigma$, we have found a sample of eight and 11 DSFGs in ALMA B6 and B7, respectively. Among these, six are detected in both B6 and B7 and the flux density ratios between those bands are consistent with high-redshift DSFGs.  Six of our DSFGs have ALMA B3 (3\,mm) observations, where they are undetected, again compatible with high-redshift DSFGs.
	
	\item The average 870-$\mu$m flux density of our DSFGs is lower than those of the classical population of single-dish-selected SMGs, and comparable with the (stacked) flux density of LBGs at $z \sim 3$. The faintest galaxies detected in our survey would have been missed even by the deepest \emph{Herschel} extragalactic surveys, for any reasonable redshift, demonstrating the relevance of our survey for studying the faint population and for exploring the link between the extreme population of SMGs or \emph{Herschel}-selected galaxies and the more abundant population of UV-selected galaxies.
	
	\item Using our sample of DSFGs we have determined cumulative submm number counts at 870\,$\mu$m and 1.2\,mm, independently, from a sparse sampling of the astronomical sky, thus remaining relatively free of cosmic variance.  This is the first determination of faint number counts with ALMA at 870\,$\mu$m. We find that the counts are lower than previously reported by a factor of at least $2\times$ at the lowest flux densities probed by our survey.

\end{enumerate}

We plan to increase the depth of the survey and the size of the area covered, as observations of more calibrators become available. Besides the obvious advantage of exploiting the significant fraction of ALMA time that is required for calibration, our approach has several other key advantages. Multi-epoch, multi-band observations mean that our detections of DSFGs are entirely secure, with confirmation via appropriate submm spectral indices, at matched resolution, and that their redshifts will ultimately be determined via the blind detection of spectral lines.  Uniquely, our approach enables morphological studies of faint DSFGs -- a more representative population of star-forming galaxies than conventional SMGs -- in fields where self-calibration is feasible on timescales of a few seconds, meaning that baselines yielding milliarcsecond spatial resolution can be exploited, for free. 

\begin{acknowledgements}
IO and RJI acknowledge support from the European Research Council (ERC) in the form of Advanced Investigator Programme, COSMICISM, 321302.  IRS acknowledges support from STFC (ST/L00075X/1), the ERC Advanced Investigator programme, DUSTYGAL, 321334, and a Royal Society/Wolfson Merit Award. We acknowledge useful discussions with Mark Swinbank and Felix Stoehr.  The analysis of all interferometric data presented in this paper has been carried out with the Common Astronomy Software Applications \citep[{\sc casa},][]{McMullin2007ASPC..376..127M}.  We would like to offer our sincere thanks to all those who have labored long and hard to deliver and operate ALMA. ALMA is a partnership of ESO (representing its member states), NSF (USA) and NINS (Japan), together with NRC (Canada) and NSC and ASIAA (Taiwan) and KASI (Republic of Korea), in cooperation with the Republic of Chile. The Joint ALMA Observatory is operated by ESO, AUI/NRAO and NAOJ.

\end{acknowledgements}

\bibliographystyle{mn2e}

\bibliography{ioteo_biblio}

\end{document}